\documentclass[twocolumn,prl,amsmath,amssymb,superscriptaddress]{revtex4-1}
\usepackage{graphicx}                       % standard graphics handler
\usepackage{color}                          % colours for the hyperlinks
\usepackage{hyperref}                       % for hyperlinks in the PDF table of contents, citations etc
\usepackage{dsfont}
\usepackage{mathrsfs}
\usepackage{bigints}
\usepackage{bbold}
\usepackage{amsthm}
\usepackage{setspace}

\newcommand{\ket}[1]{\left\vert #1\right\rangle}
\newcommand{\bra}[1]{\left\langle #1\right\vert}

\begin{document}

\title{Quantum correlations of light and matter through environmental transitions}

\author{Jake Iles-Smith}
\email{jakeil@fotonik.dtu.dk}
\affiliation{Photon Science Institute \& School of Physics and Astronomy, The University of Manchester, Oxford Road, Manchester M13 9PL, United Kingdom}
\affiliation{Department of Photonics Engineering, DTU Fotonik, {\O}rsteds Plads, 2800 Kongens Lyngby, Denmark}
\affiliation{Controlled Quantum Dynamics Theory, Imperial College London, London SW7 2AZ, United Kingdom}
\author{Ahsan Nazir}%
\email{ahsan.nazir@manchester.ac.uk}%
\affiliation{Photon Science Institute \& School of Physics and Astronomy, The University of Manchester, Oxford Road, Manchester M13 9PL, United Kingdom}
\affiliation{Controlled Quantum Dynamics Theory, Imperial College London, London SW7 2AZ, United Kingdom}

%\author[1,2,3,*]{Jake Iles-Smith}
%\author[1,3,*]{Ahsan Nazir}
%
%\affil[1]{{Photon Science Institute \& School of Physics and Astronomy, The University of Manchester, Oxford Road, Manchester M13 9PL, UK}}
%\affil[2]{Department of Photonics Engineering, DTU Fotonik, {\O}rsteds Plads, 2800 Kongens Lyngby, Denmark}
%\affil[3]{Controlled Quantum Dynamics Theory, Imperial College London, London SW7 2AZ, UK}
%\affil[*]{jakeil@fotonik.dtu.dk, ahsan.nazir@manchester.ac.uk}
\date{\today}

%\ociscodes{270.0270, 350.4238}

%\doi{\url{http://dx.doi.org/10.1364/optica.XX.XXXXXX}}

\begin{abstract}
One aspect of solid-state photonic devices that distinguishes them from their atomic counterparts is the unavoidable interaction  between %excitonic degrees of freedom 
system excitations %within the system 
and lattice vibrations 
%the thermal environment provided by 
of the host material.
This coupling may %can 
%potentially 
lead to surprising departures in emission properties between solid-state and atomic systems. 
%In some cases this %interaction
%coupling can lead to surprising %and often counter-intuitive 
%effects on the emission properties of the system. 
Here we %reveal 
%demonstrate 
%one
predict a %particularly 
striking and important %new 
example of such an effect. %where
We show that
in solid-state cavity %QED
quantum electrodynamics, %systems, 
interactions with the host vibrational environment 
can generate quantum cavity-emitter %(light-matter) 
correlations in regimes that are %what would be  
semiclassical %regimes of
for atomic systems. %emission. 
%We show that
%in solid-state cavity QED systems, 
%quantum light-matter %(emitter-cavity) %dressed 
%%entangled states 
%correlations %may
%can be generated %in solid-state cavity QED systems 
%in what are 
%%would be expected to be 
%%otherwise 
%semiclassical regimes of atomic emission, %and probed 
%through interactions with the host %a thermal 
%vibrational 
%environment. %in solid-state cavity QED systems.  
%This is true even in regimes that can be described semiclassically in %their
%the absence of such an environment, 
%for example 
%This is particularly evident, for example, when the external excitation is weak and the emitter-cavity coupling strength is dominated by cavity losses. 
This behaviour, which can be probed experimentally through the cavity emission properties, 
heralds a failure of the semiclassical approach in the solid-state, and challenges 
the notion that coupling to a thermal bath supports 
a more classical description of the system. Furthermore, it does not rely on the spectral %specific microscopic 
details of the host %thermal 
environment under consideration and is robust to changes in temperature. %, and so should 
It should thus be of relevance to a wide variety of photonic devices.
%For this experimentally relevant case, we demonstrate  
%that bath-induced dressed state 
%transitions %lead to 
%generate emitter-cavity quantum correlations
%%asymmetries 
%%in the cavity emission properties, 
%which are absent otherwise and can be probed through the cavity emission.
%We attribute this behaviour 
%to the quantum nature of the environment 
%and its resulting sensitivity to the joint eigenstructure of the cavity and emitter. This heralds 
%a failure of the semiclassical approach, and challenges 
%the notion that coupling to a thermal bath supports 
%a more classical description of the system. 
%Bath-induced asymmetries also persist over wider regions of parameter space, 
%including the Fano and quantum strong coupling regimes.
\end{abstract}

%\setboolean{displaycopyright}{true}

%\begin{document}

\maketitle
%\thispagestyle{fancy}
%\ifthenelse{\boolean{shortarticle}}{\abscontent}{}

\section{Introduction}
The diversity of systems studied in cavity quantum electrodynamics (CQED) places the subject at the heart of many prospective 
quantum and classical technologies.
Examples include single photon sources~\cite{PhysRevLett.98.117402,Michler22122000}, ultrafast optical switches~\cite{PhysRevLett.108.227402,PhysRevLett.109.166806}, and quantum gates~\cite{PhysRevLett.75.3788,Duan:2001aa,PhysRevA.78.062336}, which 
require the development of robust, scalable, and potentially strongly coupled emitter-cavity systems.  
Though the quantum strong coupling (QSC) limit and beyond---in which the system eigenstates become light-matter entangled---have now been attained for single emitters in the 
microwave regime~\cite{Wallraff:2004aa,Niemczyk10},  
it remains 
technically demanding 
to manufacture optical cavities of sufficiently high 
quality (Q) factor to unambiguously demonstrate QSC phenomena. 
Example 
systems in which 
great strides have recently been made 
towards this goal 
include single self-assembled quantum dots (QDs) within  
optical nano- and microcavites~\cite{Reithmaier04,Yoshie04,Hennessy07,Faron08,Kasprzak10}. 
Here, small mode volumes can readily  
be obtained, resulting in potentially large 
Q-factors and cavity coupling strengths that are substantial  
in comparison to the  
emitter decay rate. In conjunction with their solid state nature, this makes QD-cavity systems excellent candidates for  
future technological applications. 

Nevertheless, 
it still remains a challenging endeavour to reach the QSC regime due to  
significant cavity losses~\cite{Hu2008,Young2011}. %, which result in %a 
The broad cavity lineshape that results 
masks contributions from higher order dressed states. %thus 
%placing 
This places 
the system in an intermediate coupling regime %where the light-matter coupling
that can be described using semiclassical techniques~\cite{hu2014saturation}, 
%in which 
thus 
neglecting all %entanglement 
quantum correlations between the cavity photonic (light) and emitter electronic (matter) degrees of freedom. 
In addition to interactions with external electromagnetic fields, many CQED systems are also in contact with their host (e.g.~thermal) environment; 
for example, in QDs  
this influence is often dominated by acoustic  
phonons~\cite{PhysRevLett.105.177402,PhysRevLett.104.017402}. 
In order to explore the effect that such couplings 
have on the 
system's optical emission, 
it is necessary to 
modify the standard quantum optical  
treatments~\cite{carmichael1998statistical}, 
which may 
lead to significant  
departures from atomic-like behaviour~\cite{PhysRevB.83.165101,1367-2630-12-11-113042,PhysRevLett.113.097401,Roy2011X,McCutcheon2013,Roy2011,Hughes2013,PhysRevB.87.081308,PhysRevB.90.035312,PhysRevB.65.235311}. 
 
Here, we demonstrate that %it is possible to generate 
quantum light-matter correlations can be generated in solid-state CQED systems 
%eigenstates 
via transitions induced by the host (vibrational) environment, in regimes of %that give rise to 
semiclassical atomic CQED emission %for atomic systems 
where such an environment is absent. 
%Here, we demonstrate that %even 
%for lossy %{\bf [and weakly-driven?]} 
%CQED systems %cavity 
%within %an otherwise 
%semiclassical regimes of atomic emission, it is still possible to generate %observe signatures of 
%%the joint 
%%emitter-cavity
%quantum light-matter %(cavity photon-QD exciton) 
%%(cavity-emitter) 
%%quantum 
%correlations in the solid-state 
%%eigenstates 
%via transitions induced by the host (vibrational) environment. 
%Specifically, we explore the effect of a thermal bath  
%on the emission properties of a CQED system in several important parameter regimes, corresponding to the semiclassical~\cite{Hu2008,Young2011}, Fano~\cite{RevModPhys.82.2257,PhysRevA.82.043845}, and  
%QSC limits~\cite{RevModPhys.73.565,Bishop:2009aa}.  
Specifically, we  
show that 
the presence %quantum mechanical nature 
of the host  
environment results in optical emission that is 
observably sensitive to the joint eigenstructure of the cavity and emitter, even when the equivalent atomic %optical 
transitions %(i.e.~in the absence of the vibrational environment) 
are not, %due to cavity losses, %We also 
and quantify the resulting deviations from the semiclassical description. 
This behaviour, which may be  
probed experimentally through asymmetries in both the cavity reflectivity and emission spectra, 
also challenges the notion that the addition of a thermal environment  
should simply decohere our system 
to a more classical effective description. 
%The physics we describe is distinct %se effects are distinct %contrast markedly to the
%from %that captured by a 
%the semiclassical %Fermi Golden Rule 
%treatment of phonon-induced relaxation in cavity polariton systems~\cite{cavitypolariton}, in which %any 
%light-matter quantum correlations are inherent to the polaritonic state itself, rather than being generated by the host thermal environment {\bf [IS THIS SENTENCE NECESSARY?]}. %as is the case here. %in our present case.
%~\footnote{It is worth noting that the physics we describe is distinct %contrast markedly to the
%from that captured by a semiclassical %Fermi Golden Rule 
%treatment of phonon-induced relaxation in cavity polariton systems~\cite{cavitypolariton}, in which %any 
%light-matter quantum correlations are inherent to the polaritonic state itself, rather than being generated by the thermal environment as in our present case.}.
%regardless of whether phonons are present or not. 
%In the cases considered in this work, light-matter quantum correlations are generally absent unless the thermal bath is present.
We stress that the quantum correlations we shall describe, though mediated by the solid-state environment, are shared between the light and matter degrees of freedom of the CQED system itself (i.e.~cavity photons and internal emitter electronic states) and {\it not} 
with the host environment. 
%between the host environment and the CQED system states.

%{\bf Model ---}	
%{\bf Results ---}
\section{Model}	
\begin{figure}
\centering
	\includegraphics[width=7.5cm]{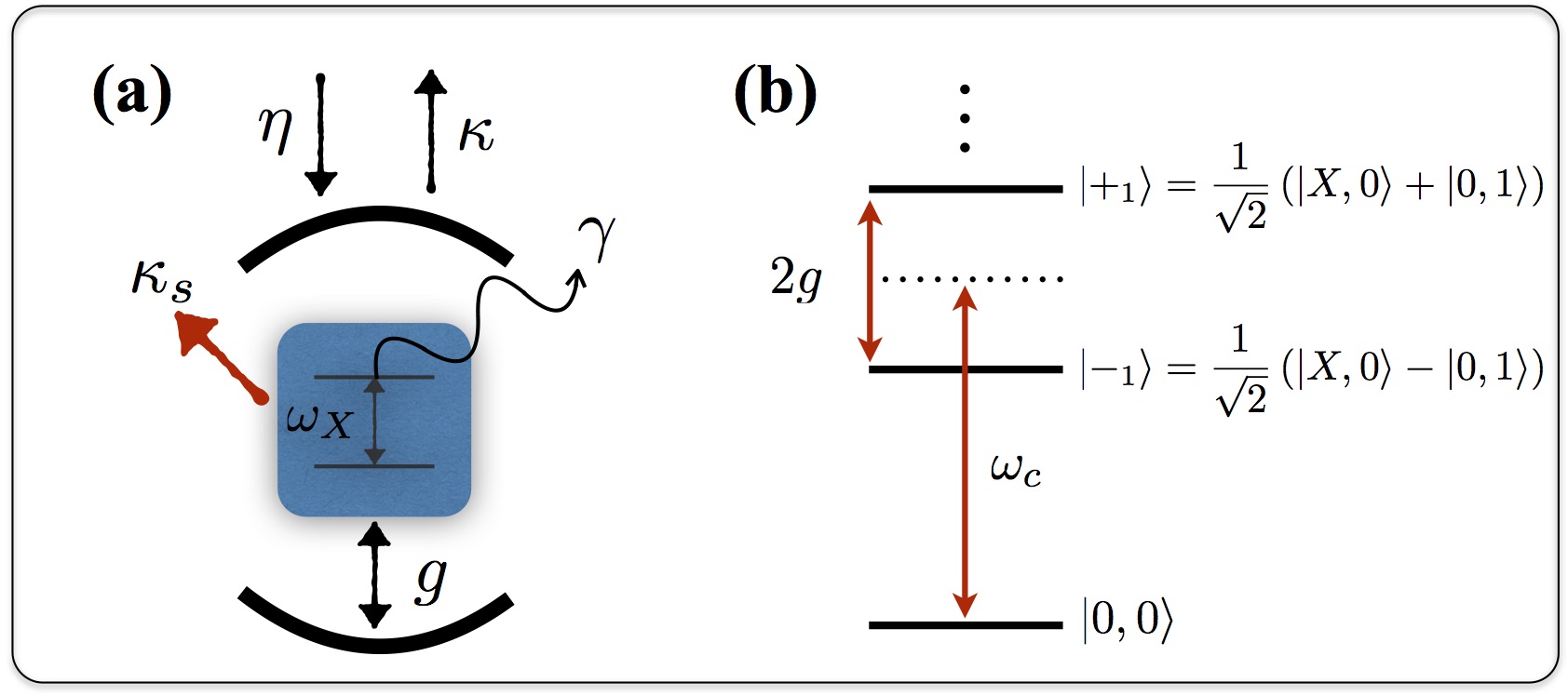}
	\caption{{(a)} Schematic of the emitter-cavity setup considered. The cavity is one sided, driven by a continuous-wave laser of frequency $\omega_L$ with  
	 strength $\eta$, and loses excitation through the top (sides) with rate $\kappa$ ( $\kappa_s$). %, and the sides with rate $\kappa_s$. 
	 The TLE decay rate is $\gamma$, and the TLE-cavity coupling strength is $g$. {(b)} The first rung of the dressed state ladder, i.e.~the lowest eigenstates of the coupled TLE-cavity system,  
	in the absence of dissipation and driving.}
\label{fig:cavity}
\vspace{-0.25cm}
\end{figure}

We  
consider a 
driven cavity coupled to a single two level emitter (TLE), shown schematically in Fig.~\ref{fig:cavity}{(a)}. This is a model of wide importance, though later we shall %also 
consider specific parameters relevant to QD-microcavity systems to provide experimental context~\cite{Young2011,Reithmaier04,Arakawa_90,arakawa_113,Hennessy07,PhysRevLett.95.067401}. 
Within a frame rotating at 
the laser %driving 
frequency $\omega_L$, and after a rotating-wave approximation, 
the system Hamiltonian is 
		\begin{equation}\label{eq:sys}
		%\begin{split}
			H_S=\delta\sigma^\dagger\sigma+g\left(\sigma^\dagger a+\sigma a^\dagger\right)+\eta\left(a^\dagger+a\right)+\mu a^\dagger a.
		%\end{split}		
		\end{equation} 
Here, $\mu=\omega_C-\omega_L$ and $\delta=\omega_X-\omega_L$ are, respectively, the cavity-laser and emitter-laser detunings, 
$g$ is the emitter-cavity interaction strength,  
$\eta$ 
is the  
cavity-laser 
coupling, %$\sigma^\dagger=\ket{X}\bra{0}$ and 
$\sigma=\ket{0}\bra{X}$ %are raising and
is the lowering operator for the TLE, and %$a^\dagger$ and 
$a$ %are creation and
is the annihilation operator for the cavity mode.

To highlight the qualitative changes in behaviour brought about in the solid-state by the presence of the %an explicit inclusion of the  
%thermal 
host environment,  
we formulate a rigorous model  
of its impact on the system 
%The first simply assumes a phenomenological pure dephasing  
%of the TLE coherences, 
%which may be approximated by  
%a semiclassical description in the appropriate limits. 
%We shall show that such a procedure is generally inadequate. %, regardless of whether the semiclassical limit is taken or not. 
%Instead, it is necessary to follow a second, more   
%retaining a  
%detailed 
%quantum mechanical 
%description of the environment. %  to properly understand its influence on the system dynamics. 
%This will be achieved 
through a polaron representation master equation~\cite{McCutcheon2013,1367-2630-12-11-113042,Roy2011X,Roy2011,PhysRevB.65.235311}. This can then be related to the cavity optical emission through the input-output formalism~\cite{PhysRevA.30.1386}. Here we consider 
the environment to be described by a collection of harmonic oscillators, with free Hamiltonian $H_B=\sum_k\nu_k b^\dagger_k b_k$, where $b_k^\dagger$ ($b_k$) is the creation (annihilation) operator for mode $k$. 
The TLE-environment interaction is given by $H_{I}=\sigma^\dagger\sigma\sum_kf_k(b^\dagger_k+b_k)$, 
%	\begin{equation}\label{Hint}	
%	H_{I}=\sigma^\dagger\sigma\sum_kf_k(b^\dagger_k+b_k),
%	\end{equation}
and its effect on the system may be described by the spectral density, which we take to 
have a super-Ohmic form appropriate to %bulk 
acoustic phonon processes, $J(\nu)=\sum_kf_k^2\delta(\nu-\nu_k)=\alpha\nu^3e^{-\nu^2/\Lambda^2}$, with $\alpha$ the coupling strength and $\Lambda$ a high frequency cut-off~\cite{PhysRevLett.104.017402,PhysRevLett.105.177402}. 

Applying the polaron transformation, $\mathcal{U}=\exp\{-\sigma^\dagger\sigma\sum_kf_k(b_k^\dagger-b_k)/\nu_k\}$, to the full 
Hamiltonian $H=H_S+H_B+H_I$ allows us to derive a master equation valid beyond the weak TLE-environment coupling regime~\cite{McCutcheon11}. 
This unitary  
generates a displaced representation of the thermal bath, removing the linear coupling term $H_I$ %of Eq.~(\ref{Hint}) 
to give 	
	\begin{align}\label{eq:polham}
	%\begin{split}
		\tilde{H}=\mathcal{U}^{\dagger}H\mathcal{U}=&
		\delta\sigma^\dagger\sigma +gB\left(\sigma^\dagger a+\sigma a^\dagger\right)+\eta(a^\dagger+a)+\mu a^\dagger a\nonumber\\
		 &+\left(XB_X+YB_Y\right) + \sum_k\nu_kb_k^\dagger b_k,
		 %\end{split}
	\end{align}
where we have absorbed the polaron shift to the emitter frequency, $\Delta_{\rm pol}=\sum_kf_k^2/\nu_k$, into the definition of $\omega_X$.	
Here, the TLE-cavity coupling strength has been renormalised by the average displacement of the oscillator environment,  
$g\rightarrow gB$, with 
$B={\rm tr}\left( B_{\pm}\rho_{\rm th}\right)$ denoting the expectation of the displacement operators $ B_{\pm}=\exp\{\pm\sum_kf_k(b^\dagger_k-b_k)/\nu_k\}$ with respect to the thermal state 
$\rho_{\rm th}=\exp\{-\beta\sum_k\nu_k b_k^\dagger b_k\}/{\rm tr}(\exp\{-\beta\sum_k\nu_k b_k^\dagger b_k\})$ at inverse temperature $\beta=1/k_B T$. The transformed operators are 
%\begin{align}
$X=g(\sigma^\dagger a+\sigma a^\dagger)$, $Y=ig(\sigma^\dagger a-\sigma a^\dagger)$, 
$B_{X}=(1/2)(B_{+}+B_{-}-2B)$, 
and $B_{Y}=(i/2)(B_{+}-B_{-})$. 

Moving into the interaction picture with respect to the 
coupled TLE-cavity Hamiltonian in the polaron frame, $\tilde{H}_s=\delta\sigma^\dagger\sigma +gB\left(\sigma^\dagger a+\sigma a^\dagger\right)+\eta(a^\dagger+a)+\mu a^\dagger a$, we 
derive a  
master equation for their reduced state, 
$\rho(t)$, by tracing out the environment within a second-order Born-Markov approximation. In essence, this procedure may be thought of as a perturbative expansion about the parameter $g/\Lambda$~\cite{McCutcheon2013}, and is thus non-perturbative in the TLE-environment coupling strength, capturing multi-phonon processes~\cite{PhysRevB.57.347}. 
%As we shall see later, the explicit presence of the TLE-cavity eigenstates in the dissipator allows the phonon environment to drive transitions between these states.
%Crucially, as the cut-off $\Lambda$ is usually much larger than any other energy scale in the problem (not just $g$) and
Nevertheless, as the renormalised coupling term $gB\left(\sigma^\dagger a+\sigma a^\dagger\right)$ appears explicitly in the system Hamiltonian $\tilde{H}_S$, this expansion does not preclude the exploration of TLE-cavity dressed states.  
%It simply enables us to move beyond weak TLE-environment couplings by retaining 
%multi-boson processes~\cite{PhysRevB.57.347}. 

Including photon emission from both the TLE and cavity, %also 
within a Born-Markov approximation and assuming the radiation field outside the cavity to have a flat spectrum~\cite{carmichael1998statistical,roy2014photoluminescence,roy2014phonon}, our master equation takes the Schr\"odinger picture form~\cite{McCutcheon2013,PhysRevA.75.013811}:
\begin{align}\label{eq:mastereq}
		%\frac{\partial\rho(t)}{\partial t} 
		%\frac{d\rho(t)}{dt} =
		\dot{\rho}(t)=& %& 
		-i [\tilde{H}_s,\rho(t)]+\mathcal{K}_{\rm th}[\rho(t)]
		%\nonumber\\
		%&
		+\frac{\gamma}{2}\mathcal{L}_\sigma[\rho(t)]
		\nonumber\\
		&+\frac{\kappa+\kappa_s}{2}\mathcal{L}_a[\rho(t)].
	\end{align}
Here, $\mathcal{L}_x[\rho]=2x\rho x^\dagger-\{x^\dagger x,\rho\}$, $\gamma$ is the TLE spontaneous emission rate,  
and $\kappa$ 
($\kappa_s$) is the photon loss rate from the top 
(sides) of the cavity. The superoperator 
%\begin{equation}\label{kthermal}
$\mathcal{K}_{\rm th}[\rho(t)]=-([{X},\Phi_{X}\rho(t)]+[{Y},\Phi_{Y}\rho(t)]+{\rm H.c.})$,
%\end{equation}
with 
$\Phi_{X}=B^2\int_0^\infty(e^{\varphi(\tau)}+e^{-\varphi(\tau)}-1)e^{-i\tilde{H}_s\tau}{X}e^{i\tilde{H}_s\tau}d\tau$, %\\% 
$\Phi_{Y}=B^2\int_0^\infty(e^{\varphi(\tau)}-e^{-\varphi(\tau)})e^{-i\tilde{H}_s\tau}{Y}e^{i\tilde{H}_s\tau}d\tau$, % 
and $\varphi(\tau)=\int_0^\infty\nu^{-2}J(\nu)(\coth({\beta\nu}/{2})\cos\nu\tau-i\sin\nu\tau)d\nu$,
accounts for interactions with the thermal bath as just described. 
When referring  
to the standard
quantum optical master equation (QOME) used to describe atomic systems, phonon processes are absent %neglected 
such that the Hamiltonian in Eq.~(\ref{eq:mastereq}) reduces to 
that given in Eq.~(\ref{eq:sys}), i.e.~$\tilde{H}_s\rightarrow H_s$, while $\mathcal{K}_{\rm th}[\rho(t)]\rightarrow0$.

	\begin{figure}[t]
	%\center\hspace{-0.5cm}
	%\centering
	\includegraphics[width=0.494\columnwidth]{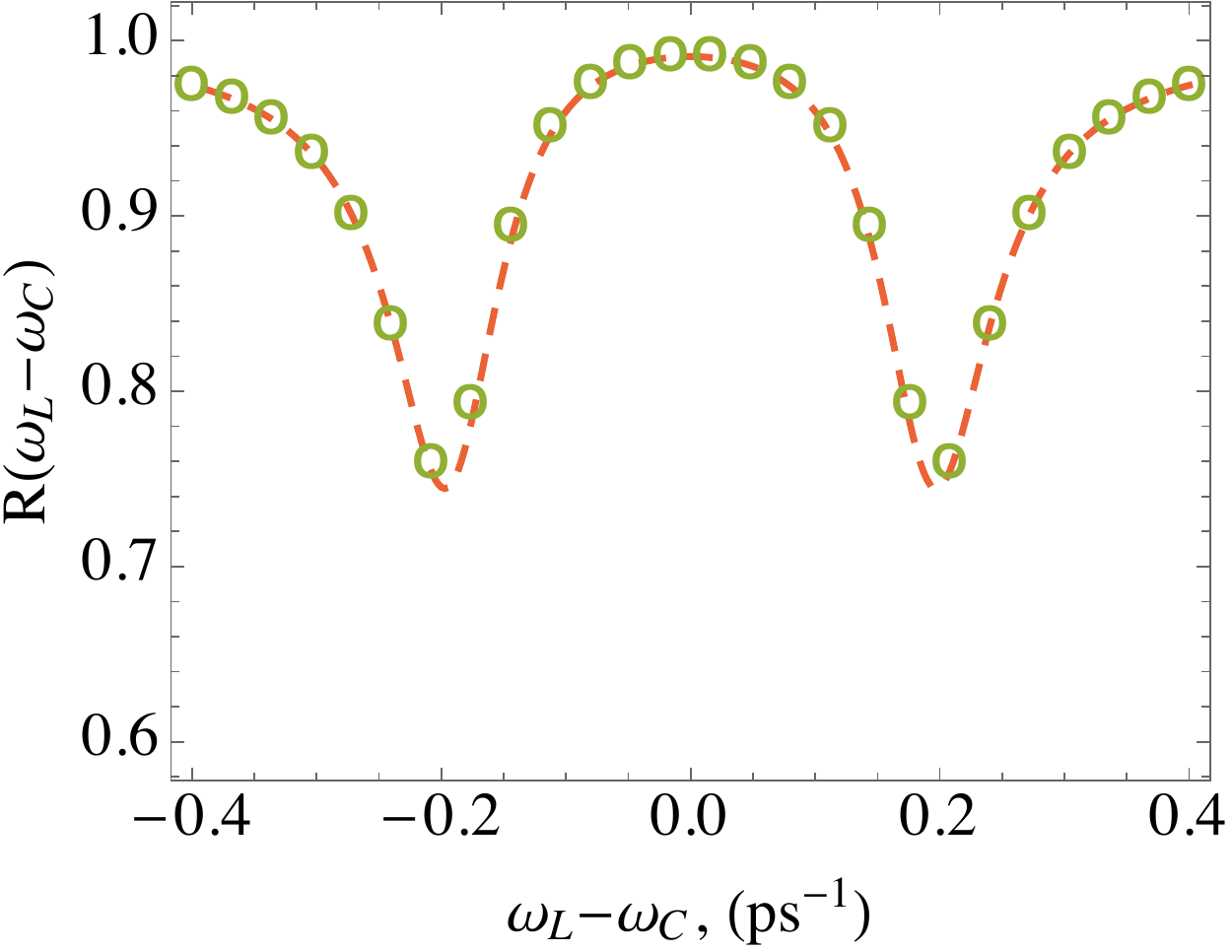}
	\includegraphics[width=0.494\columnwidth]{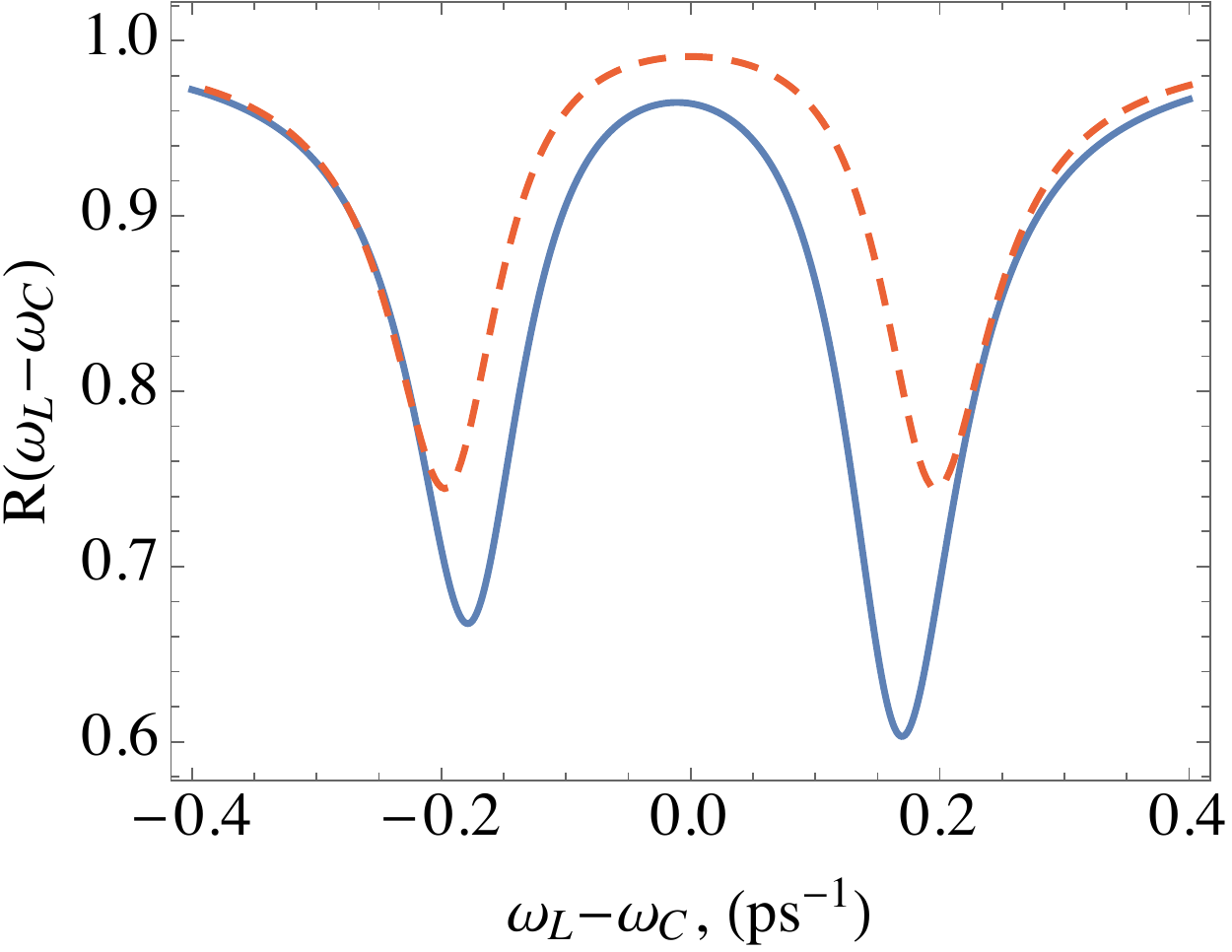}
			\caption{Steady-state reflectivity in the intermediate coupling regime, showing (left) agreement between the semiclassical theory (points) and the atomic QOME (dashed), and (right) deviations from the atomic QOME (dashed) once the solid-state environment is included %between the solid-state polaron master equation 
			(solid). %, and phase shift (d)
%in the intermediate coupling regime. %Here, the driving strength is weak, $\eta=0.01\kappa$, with other 
We choose parameters %are chosen to be 
relevant to QD-microcavity setups~\cite{Young2011,Reithmaier04,Arakawa_90,arakawa_113,Hennessy07,PhysRevLett.95.067401}:~$\kappa=g=0.2$~ps$^{-1}$, $\eta=0.001$~ps$^{-1}$, $\kappa_s=0.025$~ps$^{-1}$, $\gamma^{-1}=300$~ps, $\alpha=0.075$~ps$^{2}$, $\Lambda=2.2$~ps$^{-1}$, and $T=4$~K.} 
			\label{fig:semivspol}
			\vspace{-0.25cm}
	\end{figure}

%	\begin{figure}[t]
%	%\center\hspace{-0.5cm}
%	\includegraphics[width=\columnwidth]{semi_classical.pdf}
%			\caption{Steady-state TLE population (a), cavity occupation (b), cavity reflectivity (c), and phase shift (d) in the intermediate coupling regime for the semiclassical (points), pure dephasing (dashed) and full polaron (solid) master equations. Here, the driving strength is weak, $\eta=0.01$~ps$^{-1}$, with other parameters chosen to be relevant to QD-microcavity setups~\cite{Young2011}: $g=\kappa=1$~ps$^{-1}$, $\kappa_s=0.5$~ps$^{-1}$, $\gamma=0.01$~ps$^{-1}$, $\alpha=0.05$~ps$^{2}$, $\Lambda=2.8$~ps$^{-1}$, $2\Gamma=0.01$~ps$^{-1}$, and $T=4$~K.} 
%			\label{fig:semivspol}
%	\end{figure}
	
%Now that we have a theoretical description of the cavity-emitter degrees of system, we note that in most cavity-QED setups it is not the system expectation values that are directly probed, 
%but rather the cavity emission. 

To relate the internal emitter-cavity degrees of freedom described by Eq.~(\ref{eq:mastereq}) directly to observable experimental signatures, we note that in most CQED setups it is not the system expectation values that are directly probed, but rather the emitted cavity photons. For example, we can obtain cavity reflectivity spectra
%We can obtain the cavity reflectivity spectra
from Eq.~(\ref{eq:mastereq}) by way of the input-output formalism~\cite{PhysRevA.30.1386},
which draws a formal connection between system operators and those of the cavity emitted field. Heisenberg-Langevin equations are first used to define collective field operators for both the input, $\hat{a}_{in}$, and output, $\hat{a}_{out}$ through the top of the cavity, leading to the famous input-output relation~\cite{PhysRevA.30.1386}:
$
\hat{a}_{out}(t)-\hat{a}_{in}(t)=\sqrt{\kappa}\hat{a}(t)
$.
It is then straightforward to derive an expression for the complex cavity reflectivity in the steady state:
$
		r = \langle \hat{a}_{out}\rangle/\langle \hat{a}_{in}\rangle=1-i({\kappa}/{\eta})\langle \hat{a}\rangle
$,
where $R=\vert r\vert$ gives the reflectivity coefficient. 
%and $\phi=\arg(r)$ is the phase shift of the reflected light.
%Fig.~\ref{fig:semivspol} (left) %{(c)} and {(d)} 
%compares the cavity reflectivity spectra %and associated phase shift 
%for the QOME and semi-classical theories, demonstrating perfect agreement across the full range of detunings, highlighting the semi-classical normal mode splitting expected.

%{\bf Phonon effects in cavity spectra ---}
\section{Results}
Having outlined our theoretical approach, let us 
focus our analysis %first 
on what we shall term the intermediate coupling regime, as it has particular experimental relevance   
to %QDs in 
optical microcavities~\cite{Young2011,Hu2008} (we analyse the classical Fano and QSC limits in the Supplementary Material). 
Here, $\kappa+\kappa_s\gtrsim g>\gamma$, such that cavity leakage is significant.  
For an atomic  
system, 
this regime is characterised by a symmetric double dip structure in the steady state reflectivity spectrum on scanning a weak driving field 
through resonance. 
This can be seen by the dashed curve %points %on the left-hand-side of 
in Fig.~\ref{fig:semivspol} (left) %(a) and (b) 
which treats the TLE-cavity coupling fully quantum mechanically through the QOME. %, but includes the host environment only through a phenomenological pure dephasing process. 
%with rate $\Gamma=0.005$~ps$^{-1}$. 
The dips lie at resonances of the first two eigenstates of the TLE-cavity system, that is, the first rung of the dressed state ladder [see Fig.~\ref{fig:cavity}(b)]. 
However, unlike the true QSC regime (in which 
$g>\kappa+\kappa_s, \gamma$), the broad cavity transition obscures contributions from higher order dressed states, allowing an effective semiclassical description 
to be derived~\cite{Young2011,hu2014saturation}. 
The double dip structure may then be interpreted simply as a normal mode splitting between two classical oscillators, %with no concept of 
rather than a signature of quantum light-matter correlations. 
This can be shown in the atomic case by considering 
the relevant optical Bloch equations, obtained from the QOME: %pure dephasing master equation: 
$\dot{\langle\sigma\rangle}=-\left(i\mu+\frac{\gamma}{2}\right)\langle\sigma\rangle+ig\langle\sigma_z a\rangle$, 
$\dot{\langle a \rangle}=-\left(i\mu+\frac{\kappa+\kappa_s}{2}\right)\langle a\rangle-ig\langle\sigma \rangle-i\eta$, 
	%\begin{align}
	%\begin{split}
% 		\pd{\langle\sigma\rangle}=&-\left(i\nu+\frac{\gamma}{2}+2\Gamma\right)\langle\sigma\rangle+ig\langle\sigma_z a\rangle,\\
%		\pd{\langle a \rangle}=&-\left(i\nu+\frac{\kappa+\kappa_s}{2}\right)\langle a\rangle-ig\langle\sigma \rangle-i\eta,
	%\end{split}
	%\end{align}
with $\langle\hat{O}\rangle={\rm tr}(\hat{O}\rho)$. 
Applying a mean-field approximation 
between the cavity and TLE, such that $\langle\sigma_z a\rangle\approx\langle\sigma_z\rangle\langle a\rangle$, neglects any quantum correlations accumulated between them, 
%or in other words, the cavity and TLE 
i.e.~they remain in a product state. 
In the weak driving limit, we may further assume that on average the TLE remains close to its ground state, such that $\langle\sigma_z \rangle\approx-1$~\cite{hu2014saturation,PhysRevA.82.043845}. 
For a sufficiently lossy cavity, the resulting semiclassical theory agrees perfectly with the atomic QOME, as is demonstrated by the points in Fig.~\ref{fig:semivspol} (left). 

%for the same intermediate coupling parameters as the cavity population just described. We see that while the semiclassical theory once again agrees with the pure dephasing  
%master equation, capturing the characteristic normal mode splitting as expected, 
%in the polaron treatment 
%the suppression of the cavity %and TLE 
%population for $\omega_L-\omega_c>0$ is exhibited also in asymmetries within the cavity reflectivity. %and phase shift.

Discrepancies in the cavity reflectivity become apparent, however, when comparing %the semiclassical and QOME 
to the full (polaron) master equation relevant to solid-state CQED. %which treats {\it both} the TLE-cavity coupling and the thermal environment quantum mechanically. 
We now see a shift in the dip positions due to bath renormalisation of the TLE-cavity coupling, and an asymmetry in 
the %TLE and 
cavity reflectivity that was entirely absent in either the QOME or semiclassical calculations. 
%Clearly, such simplified approaches are unable to properly capture the physics of our system  
%in the presence of the host environment. 
Importantly, this implies that 
the semiclassical description breaks down here even within a weakly-driven and lossy cavity regime, as we shall quantify below, and thus the addition of a thermal environment \emph{generates quantum correlations} within our system. In the Supplementary Material we also show that these features cannot be reproduced by a phenomenological pure-dephasing description of the host environment, and that differences persist between the atomic and solid-state cases even at stronger driving.

%To relate population asymmetries 
%directly to observable experimental signatures 

In fact, we can attribute these asymmetric features to the quantum mechanical nature of the host environment, which plays a vital role in determining the system dynamics. 
In Fig.~\ref{fig:semivspol}, where we have chosen parameters relevant to the acoustic phonon environment common in QD-microcavity systems, 
bath-induced transitions occur on a faster timescale than other dissipative %coherence destroying 
processes, i.e.~cavity leakage and spontaneous emission. The phonon bath is 
thus sensitive to coherence shared between the cavity and QD emitter, with the result that it mediates 
transitions directly 
between the emitter-cavity dressed states. 
Specifically, when we tune the driving field to the upper dressed state resonance ($\omega_L-\omega_C\approx0.2$~ps$^{-1}$), phonon emission allows population to transfer  
from the upper to the lower dressed state,
with an associated loss of energy to the environment. 
This leads to a 
suppression of the upper dressed state population and also of the reflected light. Provided that the temperature is not too high, the inverse process, which raises population from the lower to the upper dressed state by 
phonon absorption,  
is comparatively weaker. 
%Thus, in Fig.~\ref{fig:semivspol}, we see that for negative detunings between the cavity and driving field ($\omega_L-\omega_c<0$), the lineshapes of the polaron and semiclassical theories are very similar. 
The resulting asymmetries herald a failure of the semiclassical theory, which by definition cannot be sensitive to the coherence shared between the TLE and cavity. 
This is a somewhat counterintuitive point. Naively, one might think of phonons purely as a source of decoherence, that is, as giving rise to processes that should push the system towards a more classical description. 
However, here we see that the sensitivity of the host environment to quantum correlations in fact results in the breakdown of the semiclassical description of our system, and we must instead 
reinstate a quantum mechanical explanation.

\begin{figure}[t]
%\centering
		%\includegraphics[width=0.49\textwidth]{Figures/Fig_3}
\includegraphics[height=3cm]{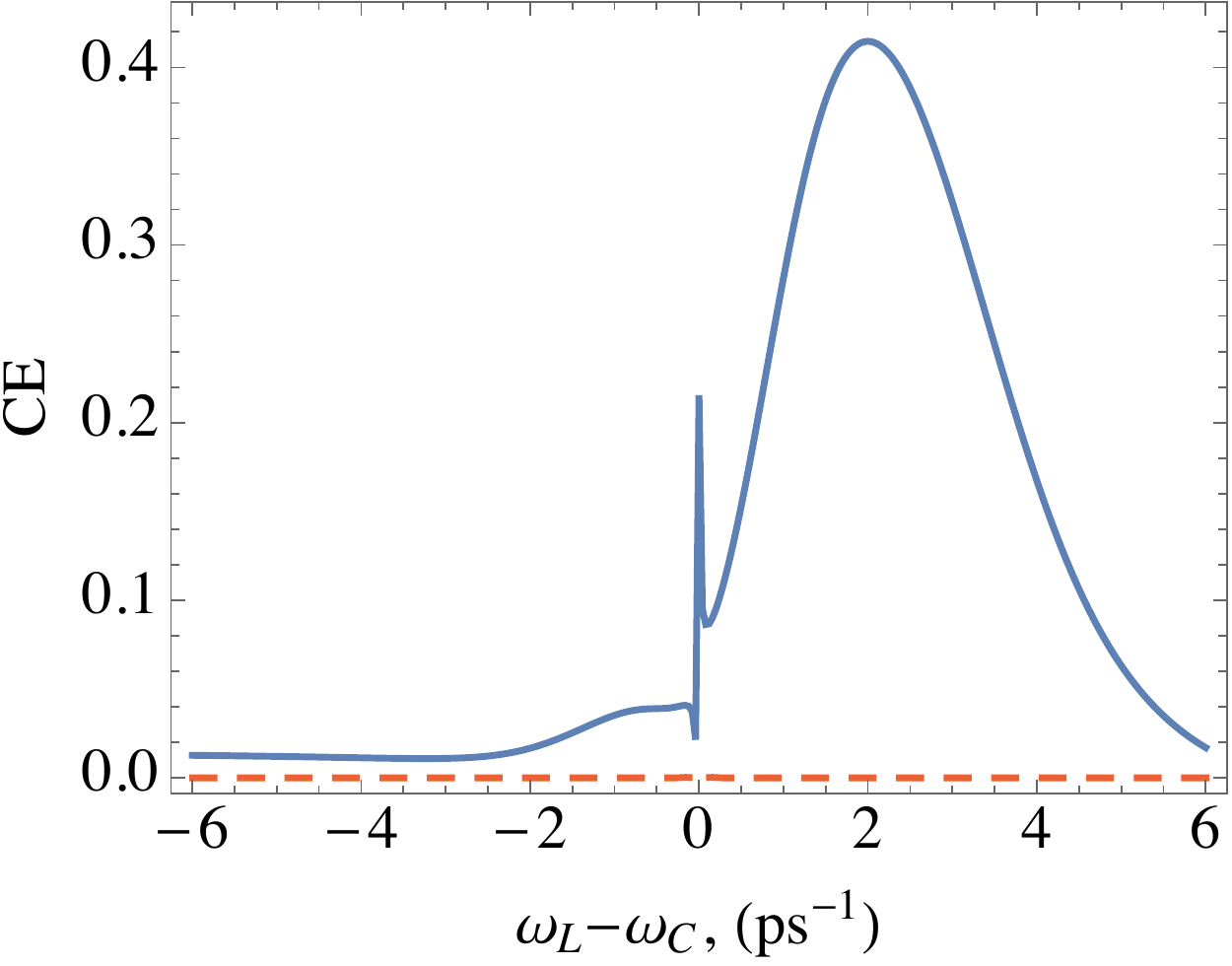}
\includegraphics[height=3cm]{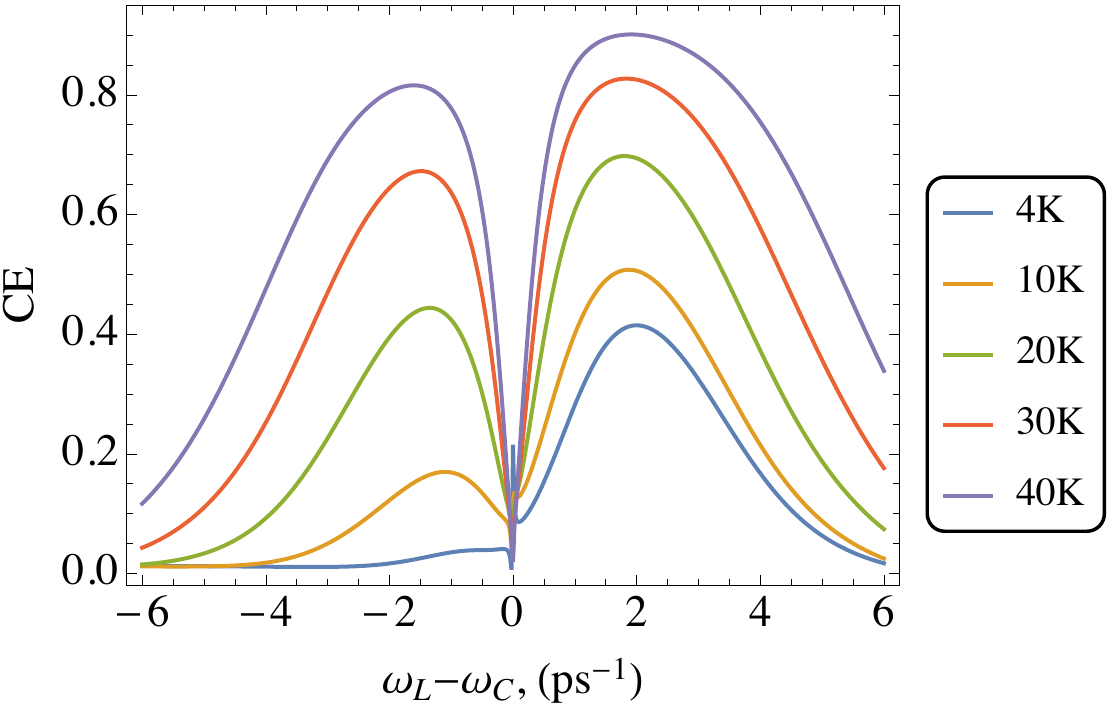}
\caption{Left:~Correlation error as a function of detuning, comparing the atomic QOME (dashed) and solid-state polaron master equation (solid) at $T=4$~K. Right:~Correlation error at increasing %over a range of 
temperature (lower to upper curves) for the solid-state master equation. All other parameters are as in Fig~\ref{fig:semivspol}.
}
		\label{fig:ce}
		\vspace{-0.28cm}
	\end{figure}

Further quantitative insight into %the deviation
departures from the semiclassical theory %due to the thermal environment %phonon interactions 
can be gained by considering the correlation error (CE), defined in Ref.~\cite{PhysRevB.85.035315} as
\begin{equation}
{\rm CE} = \vert\langle\sigma^\dagger a\rangle - \langle\sigma^\dagger\rangle\langle a\rangle\vert/\vert\langle\sigma^\dagger a\rangle\vert,
\end{equation}
which is a measure of the quantum correlations shared between the cavity photons and TLE electronic states. 
Fig.~\ref{fig:ce} (left) shows the CE for both the QOME %absence 
(dashed) and full polaron (solid) %of the  interactions 
master equations at $T=4$~K. 
%For weak driving, 
Here, the QOME shows no observable 
%deviation % from %the 
%from semiclassicality 
accumulation of correlations over the full range of detunings, confirming that we are in a semiclassical regime of atomic emission.
%theory 
%ly when driving at the cavity resonance, and most importantly not at all around the TLE-cavity dressed state resonances. 
In contrast, within the solid-state theory significant correlations are apparent across a broad range of driving frequencies, induced by the action of the thermal environment.  
%This is also seen in the full theory, however, in addition there is now a significant CE
%at other driving frequencies. 
%Due to the low temperature considered, this behaviour is asymmetric about the cavity resonance. 
%Hence, 
%%Away from resonance, 
%when driving around the lower dressed state, the host %thermal 
%environment has little effect, which leads only to a small CE %no correlations are shown and
%and the semiclassical theory giving an accurate account of the TLE-cavity behaviour.
%However, at positive detuning %we see the presence of 
%significant correlations are apparent, 
This unambiguously demonstrates the breakdown of the semiclassical description of our solid-state CQED %TLE-cavity 
system % and the emergence of light-matter correlations 
due to the presence of the host environment. %in the solid-state. 
%induced by the phonon bath.
 %This further demonstrates that phonons generate quantum correlations between the cavity and QD.
%A natural question then follows; are 

It is natural to ask whether the effects we predict are robust against variations in temperature. 
The asymmetries in the cavity lineshapes presented in Fig.~\ref{fig:semivspol} do indeed decrease as a function of increasing temperature due to environmental absorption processes balancing emission.
However, from Fig.~\ref{fig:ce} (right) we see that the CE is enhanced substantially as a function of 
%robust to changes in 
temperature for both positive and negative detuning, %and increases substantially as a function of temperature for negative detuning 
%in the latter case due to absorption processes generating correlations between the emitter and cavity. 
due to the associated increase in phonon emission and absorption rates. 
Thus the semiclassical theory remains insufficient to characterise the TLE-cavity system even as temperature is increased.
%Furthermore, even though the asymmetries in the cavity emission are suppressed as a function  of temperature, correlations induced by the phonon environment remain robust at large temperature, as demonstrated in Fig.~\ref{fig:ce}(b). 
%At higher temperatures, the CE is enhanced at negative detunings, and the failure of the semi classical theory is present over a much broader range of detunings than at low temperatures.
Despite the 
%In 
fact that lineshape asymmetries decrease at high temperatures, %we can still observe 
deviations 
%To confirm that phonon transitions do indeed drive population between the 
%dressed states, as well highlighting deviations 
from the semiclassical theory %at high temperatures 
can still be observed experimentally 
%we can
by looking at the spectra of photons emitted from the cavity when driving either the lower or upper dressed state resonantly.  
%(confirming also that environmental transitions do indeed drive population between the dressed states). 
In the former case, only at very low temperatures should we expect emission centred solely around the lower dressed state, %only at low temperatures, 
while in the latter, the same bath-mediated transitions that are responsible for correlations should lead to emission from both the lower and upper dressed states at all temperatures, in stark contrast to the conventional atomic expectation. 

\begin{figure}[t]
%\centering
		%\includegraphics[width=0.49\textwidth]{Figures/Fig_3}
\includegraphics[width=0.494\columnwidth]{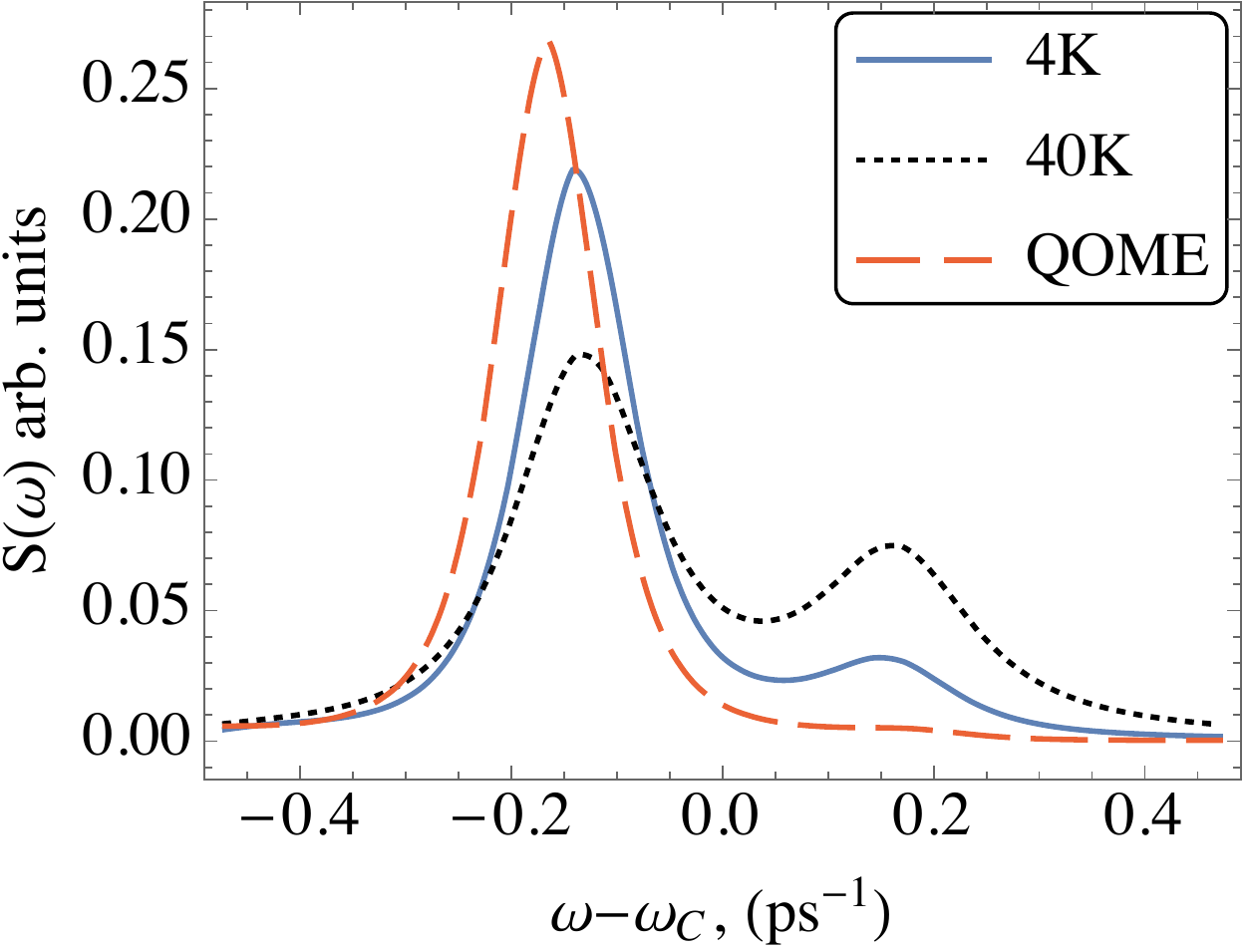}
\includegraphics[width=0.494\columnwidth]{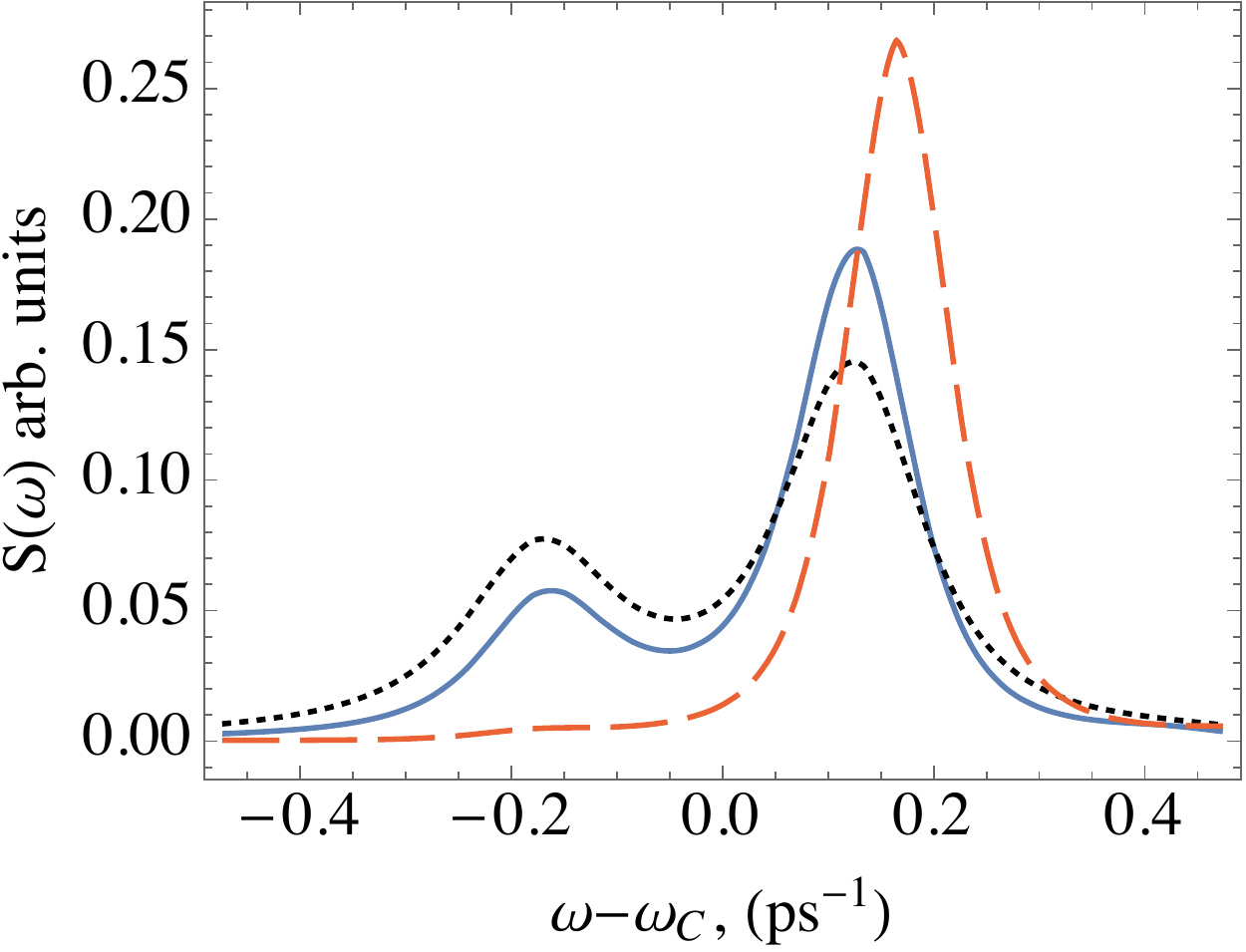}
\caption{Comparison of the cavity emission spectra using the atomic QOME (dashed) and solid-state polaron master equation (solid and dotted), under resonant excitation of the lower dressed state (left) and the upper dressed state (right) transitions. Parameters are as in Fig.~\ref{fig:semivspol}, except $\eta=0.05$~ps$^{-1}$.
}
		\label{fig:resflor}
		\vspace{-0.25cm}
	\end{figure}

In Fig.~\ref{fig:resflor} we plot  
the cavity incoherent emission spectrum, $S(\omega)\propto \operatorname{Re}[\int_0^\infty g_{inc}^{(1)}(\tau)e^{i(\omega_L-\omega)\tau}d\tau]$,  
with $g^{(1)}_{inc}(\tau)=\lim_{t\rightarrow\infty}\langle a^\dagger(t+\tau)a(t)\rangle-\vert\langle a^\dagger\rangle_{ss}\vert^2$ obtained from our master equations using the quantum regression theorem~\cite{carmichael1998statistical,mccutcheon2015optical}. 
Here $\langle a^\dagger\rangle_{ss}$ is the steady state expectation value of the cavity operator, which describes coherent emission. 
In the atomic case (dashed curves), Fig.~\ref{fig:resflor} shows the expected resonant response from a normal mode, with a single dominant peak centred around whichever is the driven dressed state transition. However, the spectra are markedly different for the solid-state CQED system, where we see a suppression of the dominant peak and the emergence of an additional feature centred around the {\it opposite} dressed state to the one being driven.
%At low temperatures (solid curve) Fig.~\ref{fig:resflor} (left) shows the expected resonant response from a normal mode, with a dominant peak around the driven lower dressed state transition.
%The spectra calculated either with the quantum-optical or polaron master equations remain similar, %largely unchanged, 
%with only a small noticeable effect of including the thermal environment in a rigorous manner. 
%However, the low temperature spectra become markedly different when we drive the upper dressed state, as shown in Fig.~\ref{fig:resflor} (right). 
%While the QOME again shows a single dominant peak, now at the upper dressed state resonance and symmetric to the previous case, in the polaron theory we see the emergence of %an additional feature.
As anticipated, this is a consequence of 
%In fact, 
%dressed state 
transitions mediated by the
host environment, %-mediated dressed state transitions 
which lead to the possibility of emission even around the undriven dressed state frequency. For lower temperatures ($4$~K), emission dominates, and the effect is thus more prominent in the right hand panel. However, at larger temperatures ($40$~K) absorption becomes almost as significant.  
%lead here to a strong suppression of the peak around the upper dressed state, and correspondingly to substantial emission at the (undriven) lower dressed state frequency. For larger temperatures %Fig.~\ref{fig:resflor}(a, 
%(dotted curves) significant population is also promoted from the lower to the upper dressed state, leading to dramatic changes in the emission spectra also when driving the lower dressed state. %where we see the emergence of a double peak structure.
Hence, by measuring the spectrum of light emitted from the cavity, we can unambiguously demonstrate the presence of bath-mediated transitions between the joint eigenstates of the emitter-cavity system %across the full range of experimentally relevant
(at both low and high temperatures), evidencing also the quantum mechanical nature of the host environment.

\section{Summary}
In summary, we have shown that the presence of a thermal environment 
allows one to generate light-matter quantum correlations %probe the dressed states of a 
in solid-state CQED systems 
within %an 
otherwise semiclassical regimes. Sensitivity of the environment to the coherence shared between the TLE and cavity leads to direct transitions between their joint eigenstates, and consequently to a breakdown of the semiclassical approach. 
The resulting experimentally observable effects persist over a broad range of parameters, %encompassing the Fano, intermediate, and strong coupling limits. They 
and should thus be applicable to a variety of CQED systems, such as QDs in micropillar and photonic crystal cavities~\cite{Young2011,Hu2008,PhysRevLett.108.227402,Reithmaier04,Yoshie04,Hennessy07,Faron08,Kasprzak10}, diamond colour centres~\cite{Faraon12,Moller12,Albrecht13}, and superconducting circuits ~\cite{Bishop:2009aa,Wallraff:2004aa,Niemczyk10}, where fluctuating resistances in the host material may be mapped to an oscillator environment~\cite{devoretleshouches}. 
%the spin-boson model{\bf[REF]}.

%{\it Acknowledgments---}
We %would like to 
thank D.~P.~S.~McCutcheon, T.~M.~Stace, K.~R.~McEnery, R.~Oulton, A.~B.~Young, and S.~Carswell for %many 
%insightful 
discussions. %conversations. 
J.I.-S. is supported by the Danish Research Councils, grant number DFF -- 4181-00416 %Engineering and Physical Sciences Research Council, 
and A.N. %is supported 
by The University of Manchester through a Photon Science Institute Research Fellowship.

%\bibliography{Phononsignaturesincavities}% Produces the bibliography via BibTeX.
\providecommand{\noopsort}[1]{}\providecommand{\singleletter}[1]{#1}%

\section{Supplemental Material}
%In this supplement 
Here we give details on several aspects of solid-state cavity quantum electrodynamic (CQED) systems supplementary to %not considered in 
the main text.
We first highlight that a pure dephasing approximation is insufficient to capture host environment induced asymmetries (and more generally environmental processes) in solid-state photonic devices. We then discuss the role of driving strength in altering the cavity reflectivity spectra and 
comparisons between the semiclassical, atomic, and solid-state cases. %asymmetries expected in cavity QED processe.
Finally, we consider the influence of a thermal %phonon induced asymmetries for 
environmental in the Fano and quantum strong coupling regimes of CQED, both of which are important to a variety of physical implementations. %of cavity QED. 
%\end{abstract}
%\setboolean{displaycopyright}{true}

%\maketitle
%\thispagestyle{fancy}
%\ifthenelse{\boolean{shortarticle}}{\abscontent}{}

\subsection{Pure dephasing noise} 

%Another approach typically used
A phenomenological approach %sometimes used 
to modelling noise processes in solid-state systems %is to 
can be obtained by assuming a pure dephasing form for the dissipator, that is, the master equation %of the form:
\begin{align}\label{eq:PDME}
\frac{\partial\rho(t)}{\partial t}=&-i\left[H_s,\rho(t)\right]+\frac{\gamma}{2}\mathcal{L}_{\sigma}\left[\rho(t)\right]+\frac{\kappa}{2}\mathcal{L}_{a}\left[\rho(t)\right]\nonumber\\&+\frac{\Gamma}{2}\mathcal{L}_{\sigma_z}\left[\rho(t)\right],
\end{align}
where the final term represents pure dephasing with rate $\Gamma$, and replaces the full environmental superoperator used in the main manuscript.  
Here, 
%the system Hamiltonian, $H_s$, is %standard driven Jaynes-Cummings Hamiltonian 
%exactly as defined in the main manuscript, and 
$\mathcal{L}_{x}[\rho(t)]=2x\rho x^\dagger-\left\{{x^\dagger}x,\rho\right\}$ denotes a non-unitary dissipator in Lindblad form,  
%(for $x=\sigma,a,\sigma_z$).
while the system Hamiltonian, $H_s$, and the other rates 
%The pure dephasing rate is given by $\Gamma$, %which controls the decay of the off-diagonal elements of the emitter density operator, 
%while the other rates %terms in the master equation 
are exactly as defined in the main manuscript. 

\begin{figure}[t!]
\center
\includegraphics[width=0.46\columnwidth]{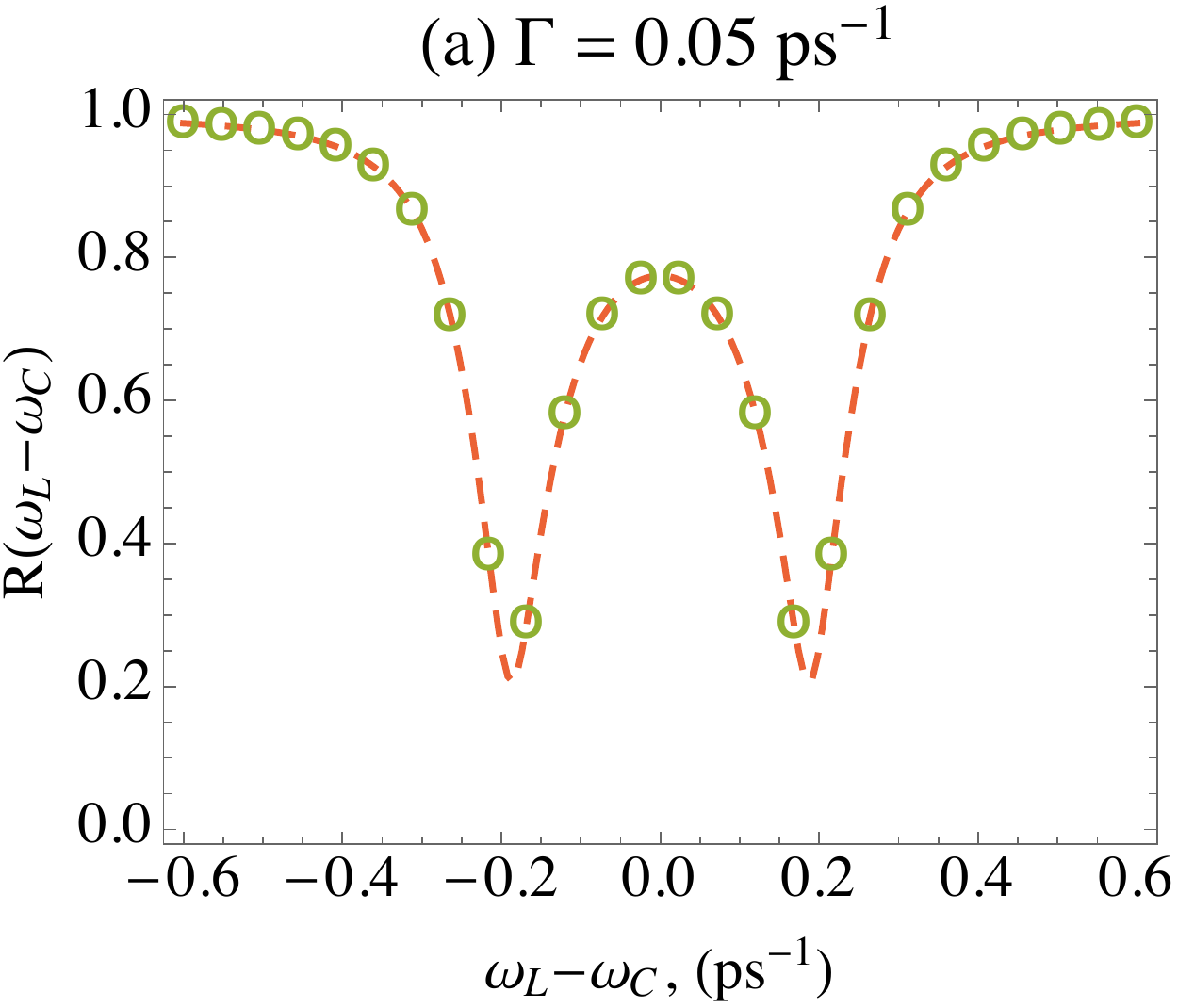}
\includegraphics[width=0.46\columnwidth]{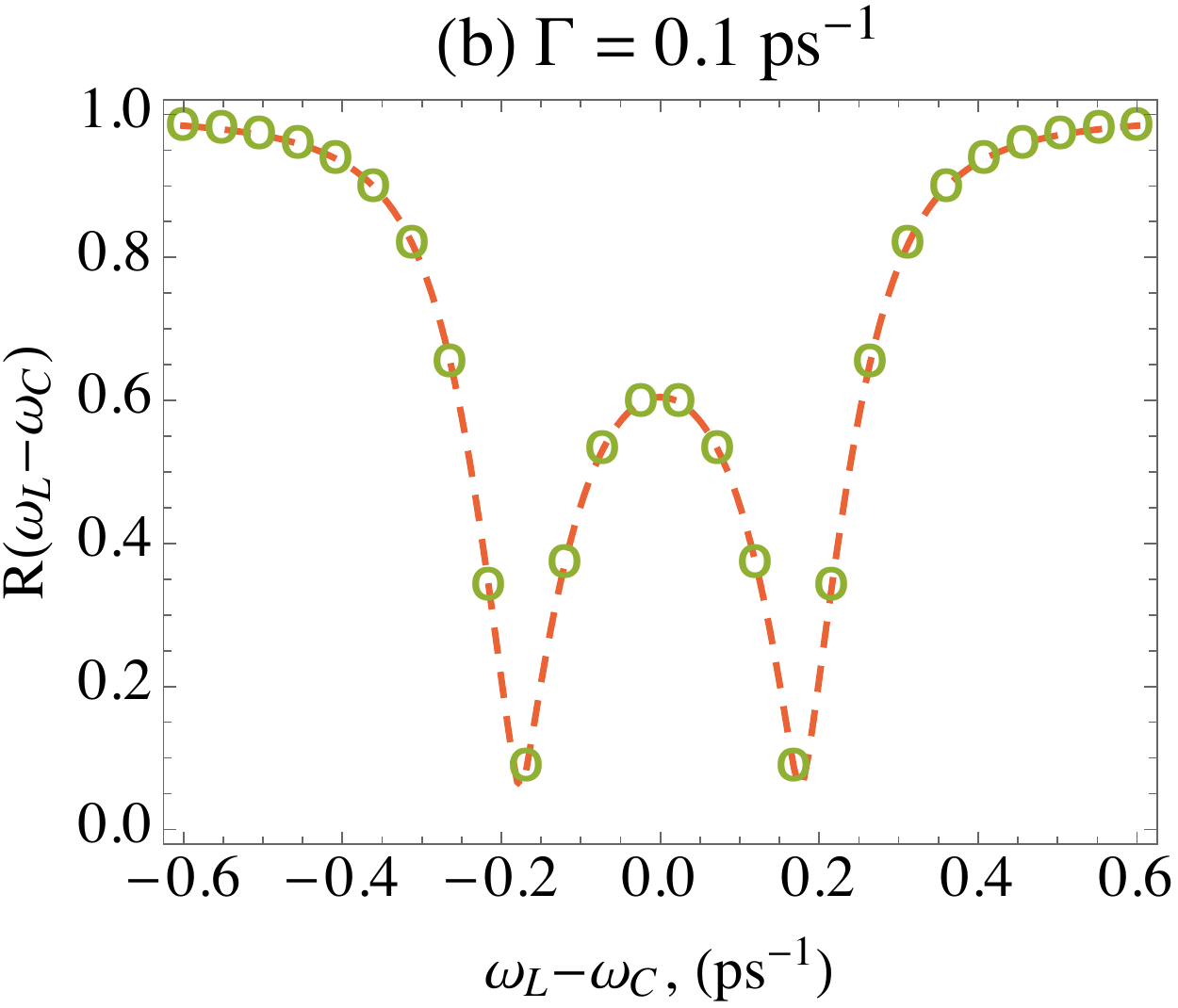}\\
\includegraphics[width=0.46\columnwidth]{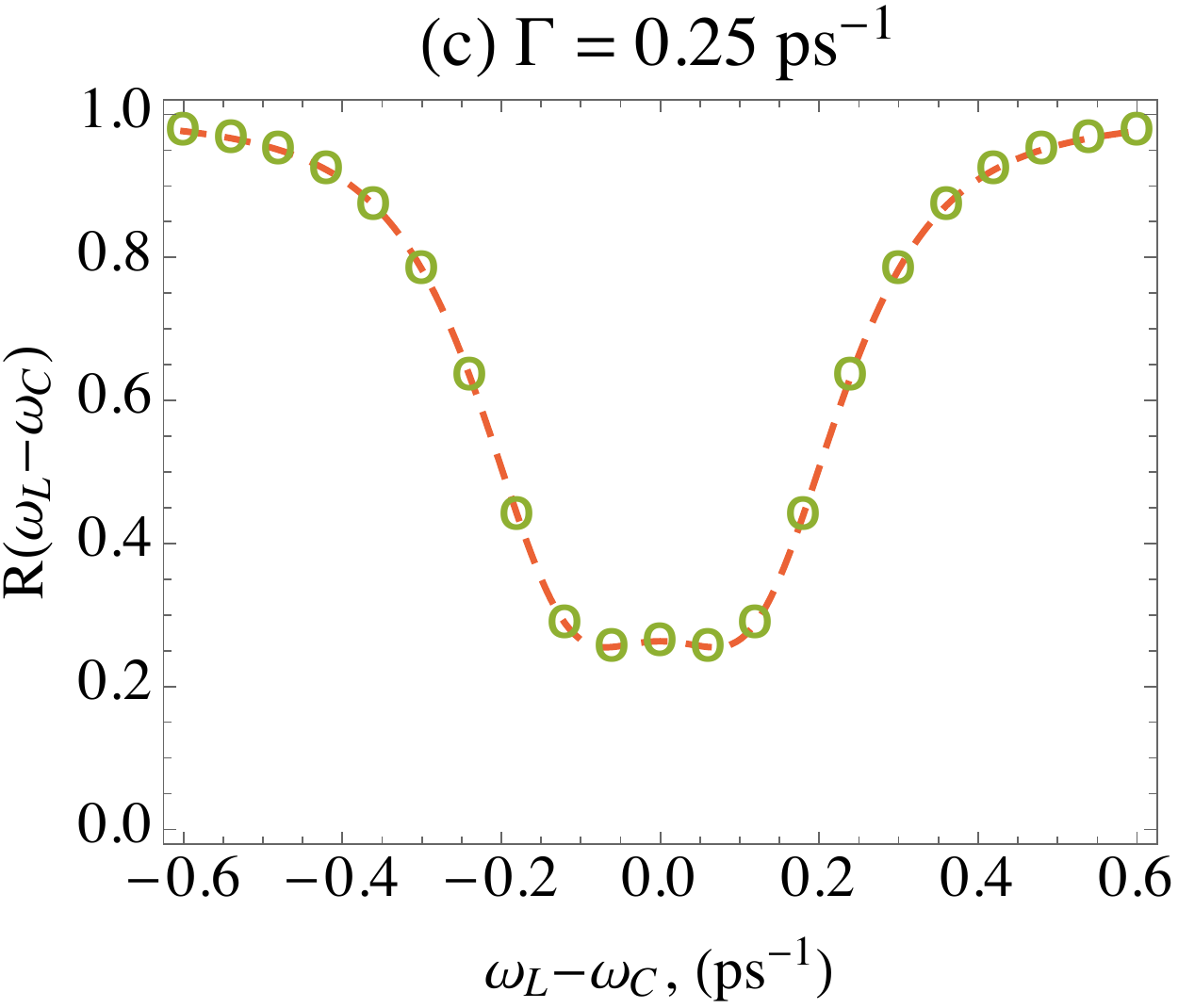}
\includegraphics[width=0.46\columnwidth]{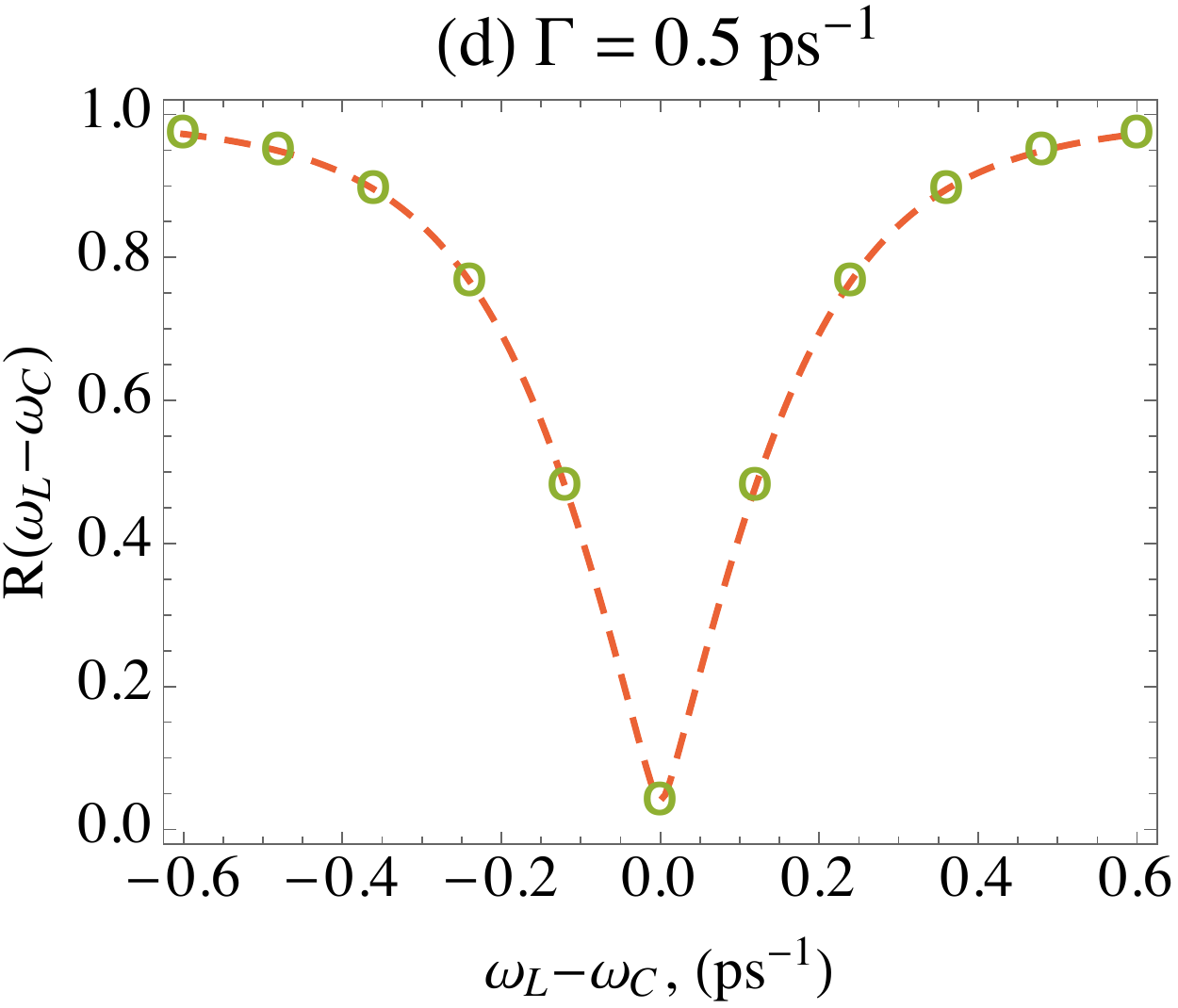}
\caption{Effects of pure dephasing on the cavity reflectivity spectra predicted by the pure dephasing master equation of Eq.~\ref{eq:PDME} (dashed) and the semiclassical theory (points). Other parameters: $\kappa=g=0.2$~ps$^{-1}$, $\kappa_s=0.025$~ps$^{-1}$, $\gamma^{-1}=300$~ps.}
\label{fig:PDeph}
\end{figure}

Despite its use in the literature, the pure dephasing master equation %has been shown to be
is insufficient to fully capture noise processes in solid-state photonic systems~\cite{nazirmccutcheonnreview}. %~\cite{McCutcheon2013}. 
One of the primary reasons for this is its disregard for the internal eigenstructure of the system under consideration, which as shown in the manuscript is essential to accurately model processes induced by the host environment. %As shown
For example, in Fig.~\ref{fig:PDeph} %for the case of 
we consider an emitter-cavity system in the same parameter regime as Fig.~2 of the main manuscript. We see that increasing the pure dephasing rate leads to a symmetric broadening of the cavity reflectivity spectra, with the semiclassical theory (now including pure dephasing) remaining valid over the full range of dephasing rates. Such a phenomenological treatment cannot, therefore, lead to the deviations from semiclassicality that we predict with a full microscopic treatment of the solid-state environment.  

%This is due to the pure dephasing leading to simply the decay of the off-diagonal elements of the density operator, unlike the full phonon theory which additionally drives transitions between the light-matter eigenstates. 

\subsection{Driving dependence and cavity asymmetry}

\begin{figure}[t!]
\center
\includegraphics[width=0.46\columnwidth]{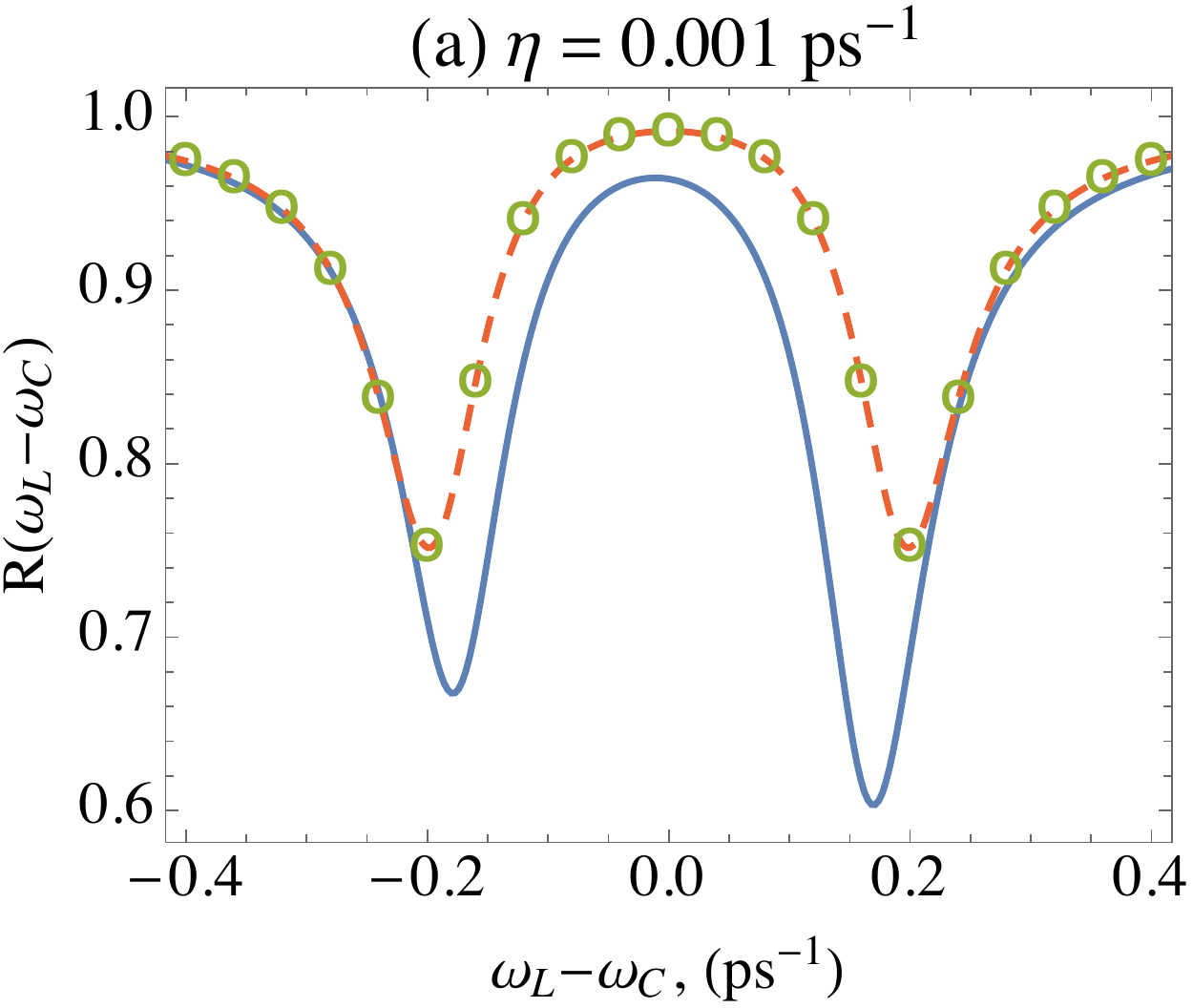}
\includegraphics[width=0.46\columnwidth]{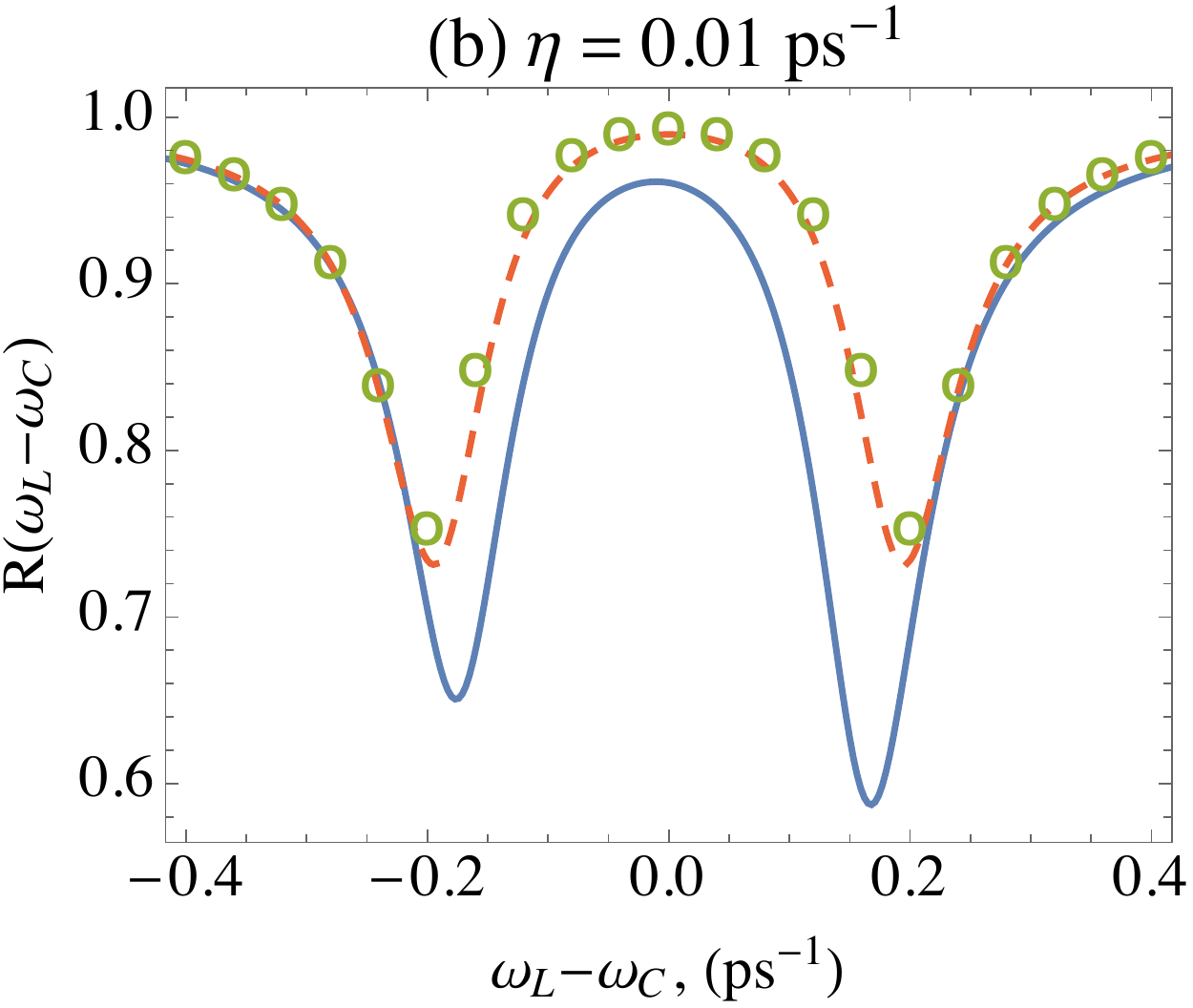}\\
\includegraphics[width=0.46\columnwidth]{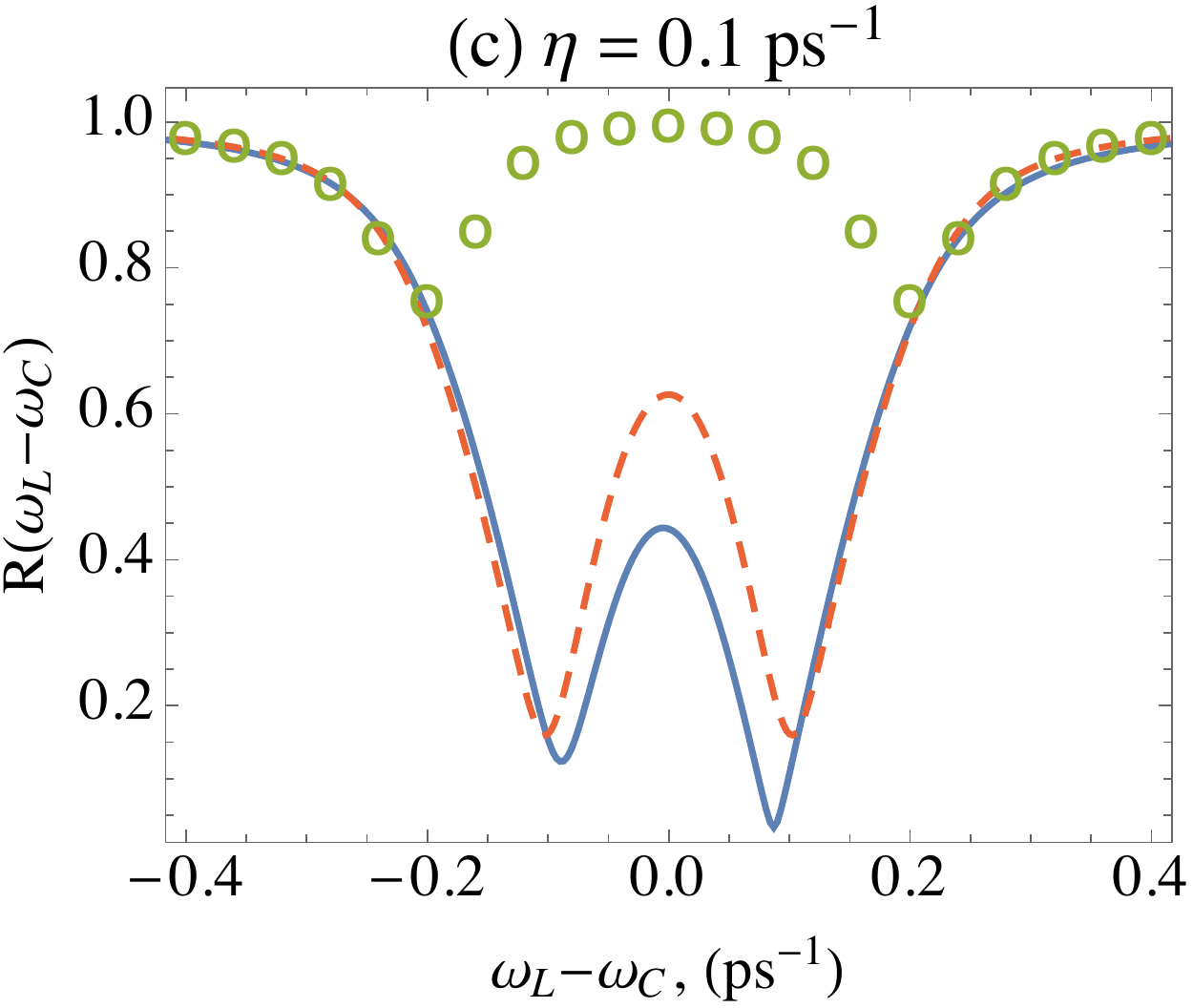}
\includegraphics[width=0.46\columnwidth]{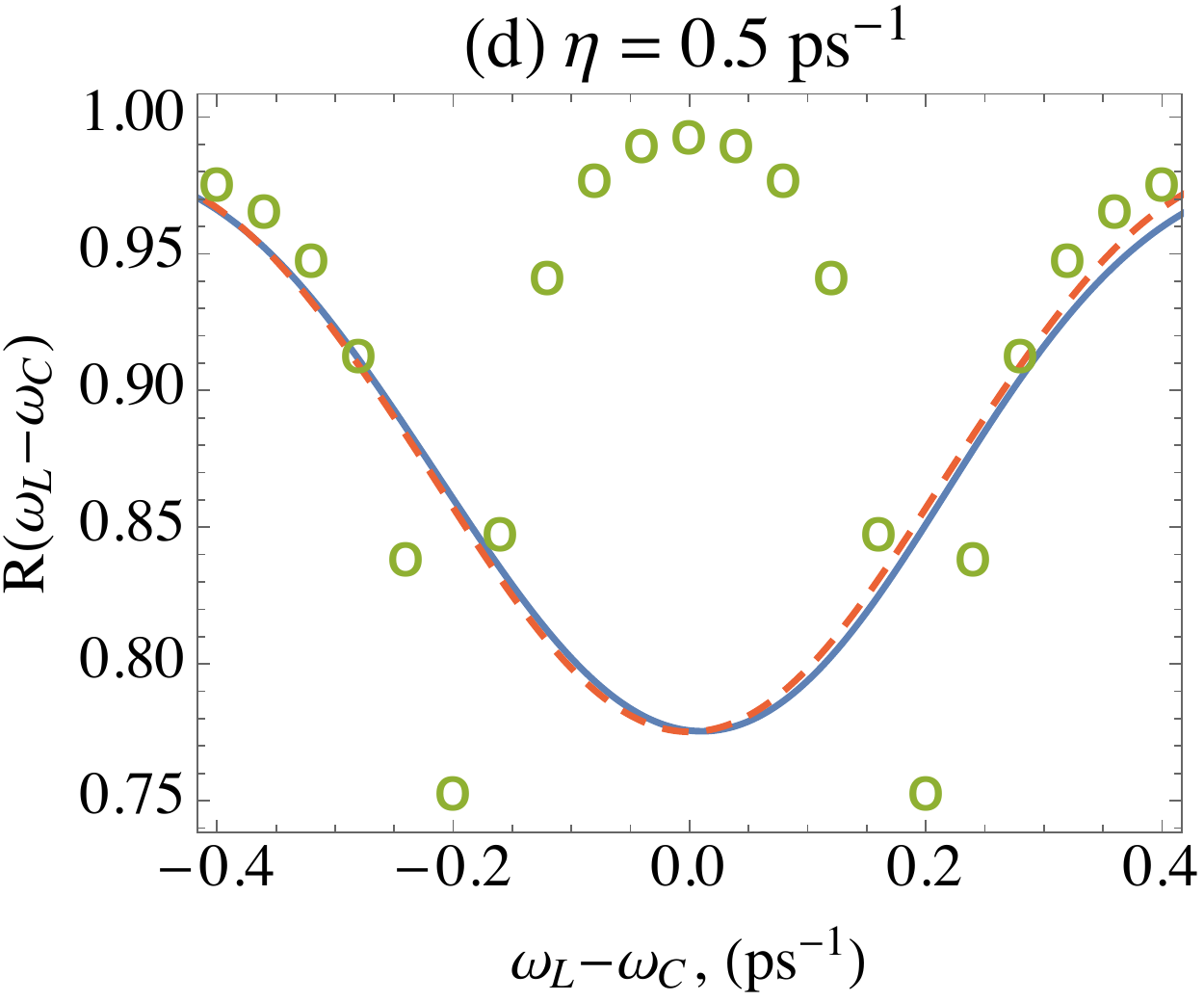}
\caption{Cavity reflectivity for increasing driving strength derived from the atomic QOME (dashed), semiclassical theory (points), and the solid-state polaron master equation (solid). Other parameters: $\kappa=g=0.2$~ps$^{-1}$, $\kappa_s=0.025$~ps$^{-1}$, $\gamma^{-1}=300$~ps, $\alpha=0.075$~ps$^{2}$, $\Lambda=2.2$~ps$^{-1}$, and $T=4$~K.}
\label{fig:figureref}
\end{figure}

The semiclassical equations derived in the main manuscript rely on the assumption of small emitter occupations, that is, $\left\langle\sigma_z\right\rangle\approx-1$.
This is valid at weak driving, when the emitter remains close to its ground state, rarely scattering a photon.
It stands to reason then that at strong driving, the semiclassical expressions derived previously will become invalid.
%This is due to the semiclassical theory being 
For example, they are unable to capture driving dependent broadening of the cavity emission~\cite{PhysRevB.91.075304}.

Despite this eventual failure, %at large driving, 
the semiclassical theory remains a robust description of the atomic (though of course not the solid-state) cavity reflectivity even when the driving strength increases by an order of magnitude from that considered in the main manuscript ($\eta=0.001$~ps$^{-1}$), as demonstrated in Fig.~\ref{fig:figureref}~(a) and (b).
As the driving strength increases yet further, the line shape broadens until the normal mode splitting is no longer visible, leaving only a single peak in the cavity reflectivity spectra as shown in Fig.~\ref{fig:figureref}~(d) at extremely strong driving. Nevertheless, signatures of phonon processes can remain observable outside the semiclassical limit, with asymmetries present even when the driving is of the same order as the cavity loss and light-matter coupling strength, see Fig.~\ref{fig:figureref}~(c).
%Interestingly, in this regime, a second semi-classical limit may be found by taking the QD transition to be saturated, that is, $\langle\sigma_z\rangle\approx0$~\cite{PhysRevB.91.075304}.
%As can be seen from the reflectivity spectra in Fig.~\ref{fig:figureref} (a-c), signatures of phonon processes remain observable across a range of driving strengths, with asymmetries present even when the driving is of the same order as the cavity loss and light-matter coupling strength.
%It is only at very large driving, when the power broadening has washed out the normal mode splitting, that phonon asymmetries are no longer observable through the cavity reflectivity. 
%Despite the semi-classical approximation failing, the phonon asymmetries persist across the driving regimes.
%Even when the double peak structure is no-longer visible at strong driving strengths, phonon processes drive transitions between the upper and lower transition inducing light matter correlations. 
%This can be seen from Fig.~\ref{fig:CE_eta}, where there remains a significant contribution to the correlation error when driving the upper dressed state transition.
%Importantly, however, predicts no light-matter correlations at these detunings regardless of the driving strength  considered.  
%  

\begin{figure}[t!]
\center
\includegraphics[width=0.92\columnwidth]{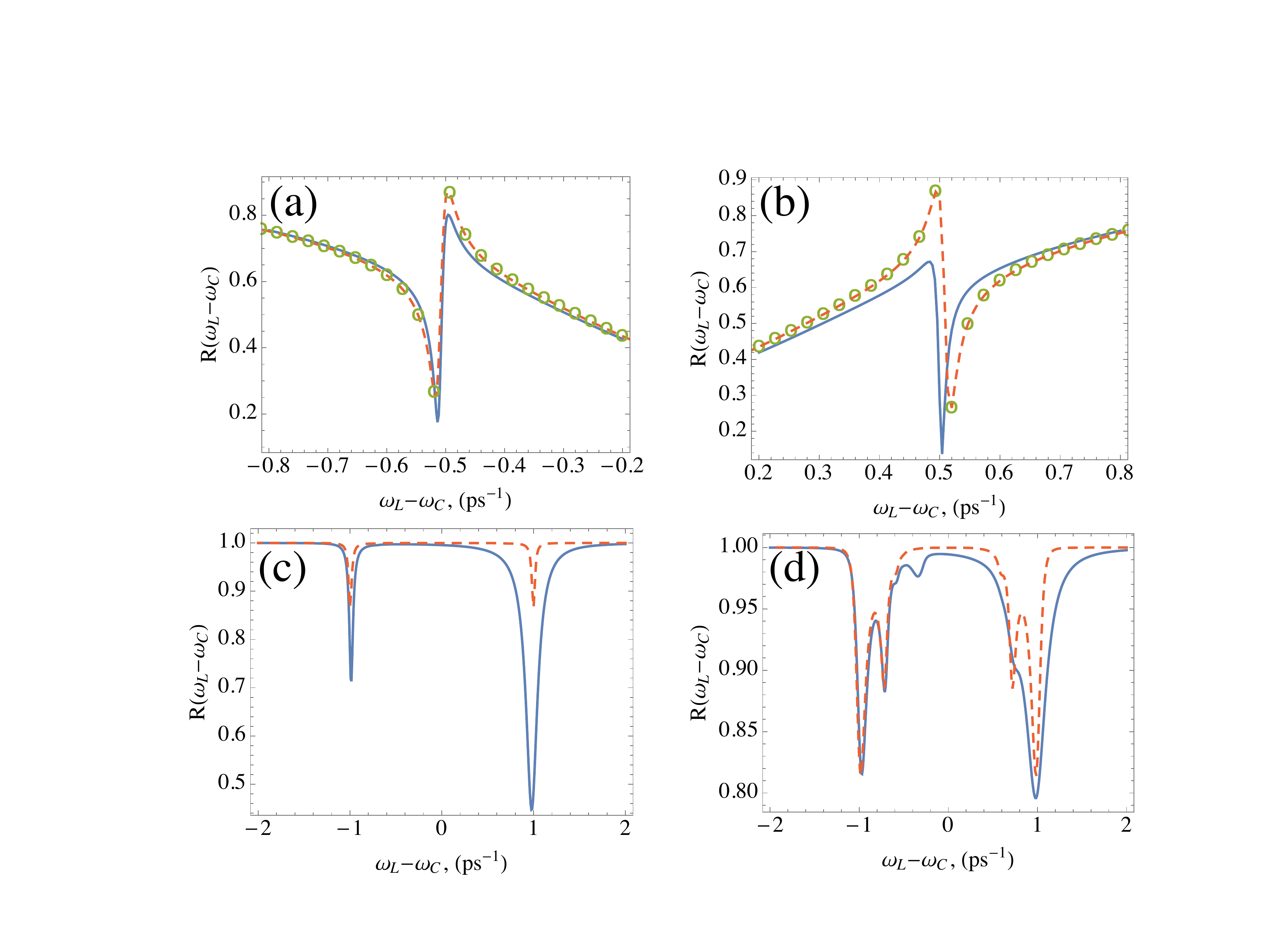}
\caption{Upper:~Asymmetry in Fano resonance profiles between emitter-cavity detuning $\omega_C-\omega_X=-0.5$~ps$^{-1}$ (a), and $\omega_C-\omega_X=0.5$~ps$^{-1}$ (b). Parameters are $\eta =0.003$ps$^{-1}$, $\kappa=1$~ps$^{-1}$, $g=0.1$~ps$^{-1}$, $\kappa_s=0.5$~ps$^{-1}$, $\gamma^{-1}= 300$~ps, $\Lambda = 2.2$~ps$^{-1}$, $\alpha=0.1$~ps$^{-1}$, and $T = 2$~K. 
Lower:~Thermal effects in the QSC regime at weak driving $\eta = 0.005$~ps$^{-1}$ (c), and strong driving $\eta=0.08$~ps$^{-1}$ (d). Parameters are $\kappa=0.1$~ps$^{-1}$, $g=1$~ps$^{-1}$, $\kappa_s=0$, $\gamma^{-1}= 300$~ps, $\Lambda = 2.2$~ps$^{-1}$, $\alpha=0.025$~ps$^{-1}$, and $T = 4$~K. Here points correspond to the semiclassical theory, dashed curves to the atomic QOME, and solid curves to the solid-state polaron theory.
}
\label{fig:fig1}
%\vspace{-0.3cm}
\end{figure}

\subsection{The Fano and quantum strong coupling regimes}

The sensitivity of the solid-state CQED system optical properties to the host environment %environment to the emitter-cavity eigenstructure 
is not restricted to the intermediate coupling regime described in the main manuscript. In fact, we shall demonstrate that similar effects emerge in the both the Fano and the quantum strong coupling (QSC) regimes.

The Fano regime of cavity QED occurs when $\kappa+\kappa_s\gg g\gg \gamma$, where $\kappa$ and  $\kappa_s$ are the cavity loss and side-leakage rates respectively.
This regime is characterised by a sharp peak in the cavity reflectivity at the emitter resonance (while the rest of the cavity line shape remains unchanged), which is the result of classical interference between two competing decay pathways~\cite{RevModPhys.82.2257}. This behaviour can be seen in line shapes given in Fig.~\ref{fig:fig1} (a) and (b). Here we see excellent agreement between the semiclassical theory and the atomic QOME, showing a sharp peak in the reflectivity spectra at the emitter resonance $\Delta=\pm0.5$~ps$^{-1}$, which is symmetric for both the positively and negatively detuned cases.
Though less pronounced than in the intermediate coupling regime, host environment induced asymmetries become apparent when comparing the cavity line shape obtained from the full polaron master equation for an emitter tuned above the cavity resonance ($\omega_C-\omega_X=0.5$) and below ($\omega_C-\omega_X=-0.5$). 

In the QSC limit [Fig.~\ref{fig:fig1} (c) and (d)], the coupling strength $g$ becomes the dominant energy scale, $g\gg\kappa+\kappa_s,\gamma$, which 
allows contributions from higher order dressed states to be resolved.
Additionally, the long cavity lifetime results in host environmental influences becoming particularly significant. This is especially true at strong driving, shown in Fig.~\ref{fig:fig1} {(d)}, where in the polaron theory the asymmetric broadening is so pronounced that it prevents us from resolving any higher order contributions to the upper dressed state resonance.

%\bibliography{Phononsignaturesincavities}% Produces the bibliography via BibTeX.
\providecommand{\noopsort}[1]{}\providecommand{\singleletter}[1]{#1}%

\end{document}